\documentclass{aastex} % submitted format

\usepackage{graphicx}
\usepackage[twocolumn]{emulateapj5}

\submitted{Accepted for publication in the Astrophysical Journal}

\newcommand{\cfour}{\ion{C}{4}}
\newcommand{\hetwo}{\ion{He}{2}}
\newcommand{\lalpha}{Ly$\alpha$}
\newcommand{\vori}{V1309~Ori} % RX0515
\newcommand{\mnhya}{MN~Hya}   % RX0929
\newcommand{\voph}{V2301~Oph} % 1H1752
\newcommand{\vaql}{V1432~Aql} % RX1940 = asynchronous

\shortauthors{SCHMIDT \& STOCKMAN}
\shorttitle{ECLIPSING MAGNETIC VARIABLES}

\begin{document}

\title{Time-Resolved {\it HST\/} Spectroscopy of Four Eclipsing Magnetic Cataclysmic Variables\altaffilmark{1,2}}

\author{Gary D. Schmidt}
\affil{Steward Observatory, The University of Arizona, Tucson, AZ 85721}
\email{gschmidt@as.arizona.edu}

\and

\author{H.S. Stockman}
\affil{Space Telescope Science Institute, 3700 San Martin Dr., Baltimore, MD
21218} \email{stockman@stsci.edu}

\begin{abstract}

Time-resolved, multi-epoch, {\it Hubble Space Telescope\/} ultraviolet eclipse
spectrophotometry is presented for the magnetic cataclysmic variables (mCVs)
\vori, \mnhya, \voph, and \vaql.  Separation of the eclipse light curves into
specific wavebands allows the multiple emission components to be
distinguished.  Accretion-heated photospheric spots are detected in \vori\ and
\voph, indicating covering factors $f<0.015$ and $\sim$0.008, and blackbody
temperatures $T_{\rm spot}>150,000$~K and $\sim$90,000~K, respectively. The
cyclotron-emitting shock on \mnhya\ is detected in the optical/near-UV and
found to occupy no more than $f=0.004$ of the stellar surface. Emission from
the accretion stream is a prominent component in \vori\ and \voph, and the
protracted eclipse ingress of UV emission lines in the latter object indicates
that the stream penetrates the white dwarf magnetosphere to a height of
$\sim$9~$R_{\rm wd}$ before it is funneled onto the magnetic poles.  The UV
emission-line spectrum of \vori\ (and to a lesser extent, \mnhya) is unusual
among mCVs, with the relative strength of \ion{N}{5} $\lambda$1240 in \vori\
matched only by the asynchronous system BY~Cam. The prominence of \ion{N}{4}
$\lambda$1718 suggests that an overabundance of nitrogen may be the most
likely explanation.

Three epochs of observation of the asynchronous \vaql\ cover about one-third
of a 50-day lap cycle between the white dwarf spin and binary orbit.  The
eclipse light curves vary enormously from epoch to epoch and as a function of
waveband.  The dereddened UV spectrum is extremely bright compared with the
optical, and the overall spectral energy distribution coupled with the
duration of eclipse ingress indicate that the dominant source of energy is a
hot ($T_{\rm wd}\sim35,000$~K) white dwarf.  This explanation also accounts
for the modest brightness variation out of eclipse and the weakness of optical
circular polarization due to cyclotron emission.  Undiminished line emission
through eclipse indicates that the eclipse itself is caused by a dense portion
of the accretion stream, not the secondary star. The temperature of the white
dwarf in \vaql\ greatly exceeds that in any other AM Her system except the
recent nova V1500 Cyg.  This, combined with its current asynchronous nature
and rapid timescale for relocking, suggests that \vaql\ underwent a nova
eruption in the past $\sim$$75-150$~yr.  The reversed sense of asynchronism,
with the primary star currently {\it spinning up\/} toward synchronism, is not
necessarily at odds with this scenario, if the rotation of the magnetic white
dwarf can couple effectively to the ejecta during the dense wind phase of the
eruption.

\end{abstract}


\altaffiltext{1}{Based on observations with the NASA/ESA {\it Hubble Space
Telescope\/} obtained at the Space Telescope Science Institute, which is
operated by the Association of Universities for Research in Astronomy, Inc.,
under contract NAS 5-26555.}

\altaffiltext{2}{Some ground-based observations reported here were obtained at
the Multiple Mirror Telescope Observatory, a facility operated jointly by the
Smithsonian Institution and the University of Arizona.}

\section{Introduction}

At least 11 of the nearly 5 dozen known AM Her variables are true eclipsing
binaries where light curves through eclipse can be used to analyze the
structure and relative importance of emission regions on the magnetic white
dwarf and in the accretion column that connects it to the mass-losing
companion.  The eclipsing systems are found with magnetic field strengths
$7<B<60$~MG (MG $=10^6$~G), span the orbital period range $1.5<P_{\rm
orb}<8.0$~hr, and include one asynchronous system.  Thus, there is reasonable
assurance that an understanding gained from the eclipsing systems can be
applied to the class as a whole (see Cropper 1990 for a general review).

Because the magnetic field funnels accreting gas onto impact regions as small
as $f=A_{\rm spot}/4\pi R_{\rm wd}^2\sim10^{-3}$, high time resolution,
$\Delta t\lesssim1$~s, plus an adequate photon rate are required to resolve the
structures.  The wavelength region of choice is the ultraviolet, where the
white dwarf and heated impact region are prominent ($T_{\rm wd}\sim15,000$~K
and $T_{\rm spot}\gtrsim30,000$~K, respectively), line and continuum emission
from the funnel gas is available, and the contribution from the late-type
secondary star is minimized. The first two papers in this series presented
{\it HST\/} time-resolved spectroscopy of DP Leo (Stockman et al. 1994,
hereafter Paper I) and UZ For (Stockman \& Schmidt 1996, hereafter Paper II);
the latter study also applied the method to ST LMi, where the accretion
footpoint undergoes ``self-eclipse'' when it rotates behind the white dwarf
limb.  DP Leo was observed in a state of very low accretion, nevertheless the
eclipse light curves revealed the contribution of an accretion-heated spot
($T_{\rm spot}\sim50,000$~K) against the surrounding white dwarf photosphere
($T_{\rm wd}\sim16,000$~K). The fractional surface area of the near-UV spot,
$f\sim0.006$, was found to be very similar to that of the optical cyclotron
source in a high state.  Observations of UZ For showed that the accretion
stream itself can be a dominant UV source in an active accretion state, owing
to both strong resonance line emission and free-bound continua. The
temperature of the hot spot was measured at $T\gtrsim30,000$~K and found to
occupy at least $f=0.01$ of the white dwarf surface area.  Attempts to model
the line emission led to a two-phase composition for the gas stream,
consisting of low filling-factor, dense clumps threading a more rarefied
flow.  Many of the observed properties of UZ For in the UV are shared by the
well-studied HU Aqr (e.g., Schwope 1998), and similar conclusions have been
reached.

In this paper we present Faint Object Spectrograph (FOS) spectroscopy of the
final 4 eclipsing magnetic cataclysmic variables (mCVs) observed in this
program.  All four are relatively bright when actively accreting, $V<17$, and
together display an interesting variety of properties: \vori\ is the
longest-period AM Her system known, with an eclipse persisting for more than
40~min.  Given the rather modest magnetic field of $B\sim60$~MG, spin-orbit
locking in this binary challenges current theories of the magnetic interaction
mechanism.  In the high state, \mnhya\ is a classic 2-pole accretion system
whose strong optical polarization enables detailed models. The weakest
magnetic field measured on any AM Her system is found on \voph, a
characteristic which has frustrated polarimetrists but motivated a number of
photometric and spectroscopic analyses.  \vaql\ is the only one of the 4 for
which an eclipse by the secondary star is in doubt; it is also the sole
asynchronous eclipsing system, made all the more unusual because the spin
period appears to {\it exceed\/} the orbital period by 0.3\%.  These and other
properties of the systems are summarized together with basic references to
previous work in Table~1.

Recently, several sophisticated techniques used to study the emission of disk
CVs have found their way to magnetic systems.  These include Doppler
tomography (e.g., Schwope et al. 1999) and maximum entropy methods (MEM) for
modeling eclipse light curves in broad bands (Harrop-Allin, Hakala, \& Cropper
1999b; Harrop-Allin et al. 1999a) as well as of full-orbit phase-resolved
emission-line spectroscopy (Sohl \& Wynn 1999; Vrielmann \& Schwope 1999; Kube,
G\"ansicke, \& Beuermann 2000). Several of these approaches utilize Genetic
Algorithms (e.g., Hakala 1995) to efficiently approach global optimization of
the many model variables.

Each of the methods has its strengths and limitations.  Doppler tomography
maps the velocity distribution of emitting regions, not their locations, and
the technique is insensitive to motions out of the orbital plane (see, e.g.,
Schwope et al. 1999).  MEM and genetic optimization codes were developed in an
attempt to deal with the tremendous ambiguity of generalized model-fitting
approaches, where the number of emitting sources may exceed the number of
independent data points. The various tools are in many ways complementary and
together represent the state-of-the-art in analytical techniques for
time-resolved studies of accretion binaries.

We have chosen to apply a more qualitative approach to interpret the eclipse
light curves of these mCVs for several reasons.  First, the photon rate in the
near-UV from {\it HST\/}+FOS cannot compete with what is now commonplace from
modern instrumentation and large ground-based telescopes working in the visible
($\sim$$0.35-1.0~\mu$m).  This, plus the low spectral resolution of FOS
effectively rule out quantitative Doppler studies of the emission lines. Time
allocation considerations on {\it HST\/} limited the observing program on each
source to windows just encompassing eclipse, and for some targets scheduling
constraints precluded obtaining both ingress and egress at the same epoch.
With such limitations, the data presented here are amenable to identifying the
location and relative importance of principle sources of line and continuum
emission, but not for unraveling their detailed structure or kinematics.

\section{Observations}

Time-resolved UV spectra of \vori, \mnhya, \voph, and \vaql\ were obtained
with {\it HST\/} at several epochs between 1996 Aug. 3 and 1996 Dec. 24 (Cycle
6, post-COSTAR), as recorded in Table~2. Setup of the FOS was as follows: the
G160L grating and 0.86\arcsec\ square C-1 aperture pair to provide a spectral
coverage of $\lambda\lambda1180-2500$, 2$\times$ substepping for a resolution
of $\sim$7~\AA; RAPID readout mode with a time resolution of 0.8201~s per
spectrum; and 5$\times$ overstepping to eliminate gaps caused by dead diodes.
Data calibration was carried out as in Papers~I and II, using the {\it
calfos\/} reduction pipeline.  Observations of \vori\ and \vaql\ were obtained
at several epochs each: the former to adequately cover the long eclipse and
the latter to sample the range of eclipse behavior shown by this asynchronous
system. The eclipse interval of \voph\ was missed in the initial observation
due to a spacecraft scheduling error, and the object was revisited a month
later. Even then, egress was not adequately covered. \mnhya\ was found in a
very faint state in the first pointing but had brightened by an order of
magnitude upon a second visit 70 days later.

The low accretion state of \mnhya\ in mid-Oct. 1996 was confirmed by a
contemporaneous ground-based spectrum that showed a visual magnitude $V>20$
and no obvious emission lines. This and a number of other supporting optical
observations were obtained over the time period of the spacecraft observations.
In the cases of \vori, \mnhya, and \voph, the Blue Channel spectrograph of the
MMT provided flux-calibrated spectra covering the range
$\lambda\lambda3300-6700$. In addition, a full orbital cycle of unfiltered
photometry and circular polarimetry with a time resolution of $\sim$45~s was
acquired during an active phase of \mnhya\ on 1996 Jan. 27 (9 months prior to
the initial {\it HST\/} visit) with the Octopol polarimeter attached to the
2.3~m Bok telescope of Steward Observatory. A $3500-6500$~\AA\ eclipse light
curve of the same object with 128~ms time samples was obtained on 1996 Mar. 10
using a programmed-readout mode of the CCD in the MMT Red Channel spectrograph
(Schmidt, Weymann, \& Foltz 1989).  For this observation, the disperser was
replaced by a plane mirror and both the target and a nearby comparison star
were located in the spectrograph slit.  The light curve of the latter object
enabled a quality eclipse profile of \mnhya\ despite the presence of light
clouds. Finally, occasional spectra and circular spectropolarimetry spanning
$\sim$$\lambda\lambda4000-8000$ were obtained with the CCD Spectropolarimeter
(Schmidt, Stockman, \& Smith 1992) attached to the Steward Observatory 2.3~m
telescope. The ground-based observations are also summarized in Table~2.

\section{Results}

\subsection{Interpreting the UV Light Curves}

Light curves were prepared from the time-resolved UV spectroscopy in 4
wavelength bandpasses in a manner similar to the procedure described in
Papers~I and II. The wavebands extracted from the dispersed spectrum are: a
near-UV interval (NUV; $\lambda\lambda 1895-2505$); the far-UV continuum (FUV;
$\lambda\lambda1254-1522$ excluding \ion{Si}{4} $\lambda$1397); and the
emission lines \cfour\ $\lambda\lambda1548,1551$ + \hetwo\ $\lambda1640$,
constructed by summing a pair of continuum-subtracted bands covering
$1529-1568$~\AA\ and $1623-1655$~\AA. In addition, a ``white-light'' curve was
derived from the zero-order undispersed image.  This bandpass has an effective
width of $\sim$1900~\AA\ for these sources and central wavelength of
$\sim$3400~\AA. The light curves for \vori, \mnhya, \voph, and \vaql\ are
shown as montages plotted vs. orbital phase in Figures~$1-4$, respectively. An
approximate absolute calibration of the 3 UV channels in units of flux per
detected count per 0.82 s sample is: \cfour+\hetwo: 2.0$\times10^{-13}$
ergs cm$^{-2}$ s$^{-1}$;
FUV: 9.3$\times10^{-16}$ ergs cm$^{-2}$ s$^{-1}$ \AA$^{-1}$; %FUV: 2.5$\times10^{-13}$ ergs cm$^{-2}$ s$^{-1}$; and
NUV: 7.4$\times10^{-17}$ ergs cm$^{-2}$ s$^{-1}$ \AA$^{-1}$. %NUV: 4.5$\times10^{-14}$ ergs cm$^{-2}$ s$^{-1}$.
Note that the two continuum bands are characterized per unit wavelength.
Neither \vori\ nor \voph\ displayed discernible variations in eclipse shape
over the various epochs, in the former despite a span of more than 2 months.
For these objects, mean light curves are shown\footnote{Count rates from the
second epoch on \voph\ were scaled up by a factor 1.5 prior to averaging to
adjust for an apparent reduction in activity over the month-long interval.}.
\mnhya\ was too faint in the initial visit to provide useful data so only the
second data set is displayed. It should be noted that the quoted accuracies of
all ephemerides (Table~1) are sufficient to phase the data presented here to
an accuracy of $\sigma_\varphi<0.004$ for all but \vori, where the uncertainty
is $\pm0.008$.

The mean spectrum of each of the 4 objects as it emerged from the processing
pipeline showed evidence for the 2200~\AA\ silicate depression due to
interstellar absorption.  Using the extinction curve from Cardelli, Clayton, \&
Mathis (1989) and $R=A_V/E(B-V)=3.1$, corrections were applied to each
spectrum with an eye toward producing as smooth a continuum as possible in the
region $\sim$$1700-2500$.  The results, quoted in Table~1, indicate visual
extinctions in the range $0.1 \le A_V \le 0.7$~mag, with uncertainties of
$0.05-0.1$~mag.

For typical system parameters, eclipse by the secondary of a point on the
white dwarf has a maximum duration of $\Delta\varphi\sim\pm0.04$
($\sim$$10-40$~min total length), with a weak dependence on binary period.
Eclipse transition across the white dwarf disk itself requires a phase
interval $\Delta\varphi\gtrsim0.004$ (at least 30~s), with the inequality
reflecting the dependence on primary star mass, orbital period (greater for
shorter $P$), and inclination (greater for smaller $i$). Previous studies
(e.g. UZ For, Paper II) have found that the accretion column is a strong
source of not only line emission, but also optical-UV continuum. Because the
gas stream approaches the primary in a curved, prograde path (see, e.g., Lubow
\& Shu 1975) and dynamically couples to the magnetic field en route, covering
of the accretion column can commence prior to the white dwarf eclipse and
continue long after. Moreover, depending on the colatitude of the impact
region on the star, the stream may arc out of the orbital plane and undergo
only a partial eclipse. At the accretion shock, optical-IR cyclotron emission
typically originates over a region of the stellar surface with $f_{\rm
cyc}\sim0.001$. A somewhat larger portion of the disk $0.001 \lesssim f_{\rm
spot} \lesssim 0.1$ is heated by low-level accretion or the EUV-bright base of
the accretion column to a characteristic temperature $kT\approx10-30$~eV. For
a circular profile, eclipse of this hot spot will produce an abrupt drop in
the FUV flux over an interval 1~s $\lesssim \Delta t \lesssim 10$~s, depending
on the observed waveband. The smallest spots are not resolvable at our sampling
frequency (and measured UV count rates).

Therefore, the baseline behavior that we expect through eclipse is:
1)~Initially a gradual extinction of the emission-line flux, beginning with
the gas stream near the secondary and proceeding along the flow; 2)~A brief
decline in the white light and NUV continuum bands that is well-resolved at
the FOS sampling rate, corresponding to the limb of the secondary beginning to
pass over the white dwarf disk; 3)~An abrupt, possibly unresolved, drop in the
FUV and to a lesser extent, the NUV bands, as the accretion-heated spot on the
white dwarf is eclipsed.  This transition will very nearly coincide with a
similar extinction of the cyclotron flux in the white light channel.
Continuing through the light curve, we find  4)~The conclusion of the resolved
white dwarf eclipse; 5)~Continued obscuration of the prograde portion of the
gas stream; 6)~A period of nearly constant minimum light lasting up to several
minutes; and 7)~$-$~11) An egress sequence that is a vertically-mirrored
repeat of steps 1)~$-$~5), in order, as each component is progressively
exposed. Barring changes in viewing perspective or in the beaming factor of an
emission component, amplitudes of corresponding ingress/egress transitions
will be identical.  The Rosetta Stone for this basic behavior is UZ For (Paper
II).

\subsection{\vori\ (RX J0515.6+0105)}

\vori\ has the longest orbital period of the known AM Her binaries at 7.98~hr.
The UV (Figure~5) and optical spectra (Garnavich et al. 1994; Shafter et al.
1995) are dominated by continuum and line emission from the accretion stream.
Buckley \& Shafter (1995) report the detection of weak, variable circular
polarization in a white-light optical bandpass. Spectropolarimetric
observations of the system obtained by us in 1994 Nov. (not shown) confirm a
low level of polarization for the system ($|v|\lesssim2\%$) as well as the
existence of cyclotron emission harmonics centered near
$\lambda\lambda4425,5510,6625$ in both total flux and polarization.
Identifying these with harmonic numbers $m=5,4,3$, respectively, we
substantiate previous estimates of the primary magnetic field strength near
61~MG (Garnavich et al. 1994; Shafter et al. 1995; Harrop-Allin et al. 1997).
The polarization appears to vary synchronously on the orbital period and is
probably strongly diluted by light from the accretion stream. It is remarkable
that such a wide binary can maintain synchronism given the implied accretion
rate and modest magnetic field (see also Frank, Lasota, \& Chanmugam 1995 and
Harrop-Allin et al. 1997). Nevertheless, the 1996 UV light curves shown in
Figure~1 are so similar to published visible-light profiles from $1993-1997$
(e.g. Shafter et al. 1995), and so unlike the variable light curves exhibited
by asynchronous systems (e.g., \S3.5), that we conclude the spin and orbital
motions are indeed locked. Interestingly, \vori\ displays an extremely
peculiar X-ray behavior (Walter, Scott, \& Adams 1995; de Martino et al.
1998), accreting in bursts as short as a few seconds duration whose local
$\dot m$ values fall in the domain of buried accretion shocks (Beuermann \&
Woelk 1996), and which cool primarily in the soft X-rays (e.g., Szkody 1999).
This extreme flickering does not appear in the UV light curves recorded by FOS.

The \vori\ eclipse light curves shown in Figure~1 and at an expanded scale in
Figure~6 confirm the dominance of the accretion stream in the UV.  Apart from
inflections in the continuum bands at the beginnings of ingress and egress,
all four channels display protracted transitions ($\Delta\varphi=0.025$). The
lack of sharp transitions in the white-light continuum places an upper bound on
the cyclotron emission in this band ($\lesssim$5\%).  The inflections
themselves are of approximately the correct duration
($\Delta\varphi\sim0.005$) to ascribe to an eclipse of the white dwarf, but
this explanation requires a hot photosphere, $T_{\rm wd}>30,000$~K. The
interpretation is also complicated by the simultaneous egress of the
line-emission region at $\varphi=0.035$ as is evidenced by the emission-line
channel.

On the other hand, a sharp drop in the FUV channel at $\varphi=0.952$ is
strong evidence for the presence of an accretion-heated spot on the white
dwarf. The data are sufficiently noisy that we show this light curve in
Figure~6 after smoothing by a running mean of width 10 samples (8~s).  The
ingress time for the spot as measured from the two contacts $c_1,~c_2$
indicated in the figure is $6\pm2$~s, implying that the transition is
unresolved in the smoothed data. We therefore can set only an upper limit to
the azimuthal extension of the spot as $<$$2\times10^8$~cm. The implied
covering factor for a circular spot on a 0.6~$M_\sun$ white dwarf is $f_{\rm
spot}<0.015$, possibly larger if the star is over-massive.  The reduction in
FUV flux corresponding to this drop is $\sim$2.2 counts sample$^{-1}$, or a
dereddened flux of $3.2\times10^{-15}$ ergs cm$^{-2}$ s$^{-1}$ \AA$^{-1}$ at
an effective wavelength of $\sim$1400~\AA.  For a distance $D>500$~pc, the
corresponding blackbody temperature is $T_{\rm spot}>150,000$~K ($kT>13$~eV).

For \vori, removal of the 2200~\AA\ feature was accomplished with a minor
amount of absorption, $A_V=0.15\pm0.05$~mag, despite a distance of at least
500~pc.  The dereddened spectra are shown in Figure~5 and characterized by the
line strength measurements compiled in Table~3.  Out of eclipse, \vori\ shows
the high excitation line \ion{O}{5} $\lambda$1370 as well as extraordinarily
strong \ion{N}{5} $\lambda$1240. The lines persist weakly through eclipse
($0.98 < \varphi < 0.02$), attesting to a small, continuously-exposed portion
of the funnel or corona on the secondary star. The remarkable strength of
\ion{N}{5} $\lambda1240$ has already been noted by Szkody \& Silber (1996)
from an {\it IUE\/} spectrum.  The ratio \ion{N}{5} $\lambda1240$/\cfour\
$\lambda1549 = 7.2$ is exceeded among known mCVs only by BY~Cam at one epoch
(Bonnet-Bidaud \& Mouchet 1987; Shrader et al. 1988); indeed among mCVs in
general this ratio rarely approaches unity.  With $\sim$30~eV separating the
ionization stages, the very large ratio implies at minimum a hot ionizing
continuum and high density conditions. However, more detailed analysis by
Bonnet-Bidaud \& Mouchet found that the line ratios for BY~Cam were so extreme
that a nitrogen enhancement was indicated. Proposed enrichment mechanisms
included accretion from a companion whose hydrogen-rich layers had already
been stripped, or mixing of normal stellar material with nuclear-processed
matter: either nova ejecta or possibly surface material of the white dwarf.
For stripping of an initially main-sequence secondary, the orbital period of
\vori\ would require that well over a solar mass of material has been shed by
the secondary over the lifetime of the binary -- a very unlikely scenario.
The fact that the \ion{N}{5}/\ion{Si}{4}, \ion{N}{5}/\cfour\ line ratios
place \vori\ precisely atop the tight grouping of 4 of the 5 measurements for
BY~Cam (Bonnet-Bidaud \& Mouchet 1987) would seem to argue against a mechanism
that involves mixing of unprocessed and processed material -- either from a
previous nova eruption or from within the secondary star. Nevertheless, a
nitrogen abundance anomaly would seem to be the best explanation since
\ion{N}{4} $\lambda$1718 is also overly strong in both systems relative to the
other atomic species (Figure~5 and Table~3).

\subsection{\mnhya\ (RX J0929.1+2404)}

The interpretation of \mnhya\ is relatively straightforward.  Except for the
eclipse, circular polarization varies smoothly from $-$20\% to +10\% through
the 3.39~hr period (Buckley et al. 1998b).  Our orbit of polarimetry displayed
in Figure~7 exhibits this behavior at a time resolution of 45~s
($\Delta\varphi=0.004$). The presence of both senses of polarization during
the orbital cycle indicates a 2-pole accretion system.  Ramsay \& Wheatley
(1998; cf. Buckley et al. 1998b) have proposed a model in which a dominant,
positively-polarized pole is hidden behind the stellar limb except during the
interval $0.45 \lesssim \varphi \lesssim 0.75$.  A weaker,
negatively-polarized pole is visible between $0.65 \lesssim \varphi \lesssim
0.25$ and responsible for the strong polarization around the time of eclipse.
Though a portion of our optical light curve is complicated by the passage of
clouds, the polarization curve is unaffected and consistent with this
interpretation. Furthermore, the ${\cal S}$-wave in circular polarization in
the interval $\varphi=0.4-0.55$ can be explained in this model as the
successive disappearance of the negative pole and appearance of the positive
pole over the white dwarf limb, with the double sign change due to viewing
each of the slightly elevated shocks ``from below'' (ref. the polarization
undershoot in VV Pup; Liebert \& Stockman 1979).  A flux minimum at this time
is consistent with the grazing viewing aspect of both accretion regions.

Our single epoch of FOS spectroscopy of \mnhya\ presented in Figure~2 shows
sharp, well-defined transitions in the white-light and NUV bands, indicating a
compact cyclotron-emitting shock.  Very little flux is detected in the two
high-frequency channels. The light curves closely resemble optical eclipse
profiles measured by Buckley et al. (1998a) but are quite unlike the X-ray
light curve (Buckley et al. 1998b), where a broadened eclipse and hardening of
the spectrum indicates absorption by the intervening accretion stream.  In
Figure~8 the FOS white-light and NUV curves are compared to an eclipse profile
obtained on 1996 Mar. 10 with the MMT using a CCD continuously clocked at a
rate of 7.8 rows s$^{-1}$.  In the latter curve, the effective time resolution
is set by the FWHM of the seeing profile at $\Delta t\sim0.7$~s, similar to the
FOS data. Ingress and egress of the bright spot are measured from the optical
light curves as $\Delta t(c_1,c_2)=2$~s, and $\Delta t(c_3,c_4)=3$~s, i.e.
essentially unresolved.  The implied linear dimension on the white dwarf is
$\lesssim$$1\times10^8$~cm ($f_{\rm cyc}\lesssim0.004$).

A non-zero flux and concave profile through minimum light is common to at
least the white-light and NUV channels.  The UV spectrum from the eclipse
interval, $0.987<\varphi<0.024$, shown in Figure~9, exhibits strong emission
lines superposed on a weak, rather red, continuum.  Clearly, a small portion
of the stream remains exposed even at mid-eclipse.  The extinction correction
for \mnhya\ was based on the out-of-eclipse spectrum and amounts to only
$A_V=0.10\pm0.05$~mag. It is important to note that, like \vori, \mnhya\ is
characterized by an unusually strong \ion{N}{5} $\lambda1240$ line, with the
flux ratio \ion{N}{5} $\lambda1240$/\cfour\ $\lambda1549=3.0$.  Once again,
\ion{N}{4} $\lambda$1718 appears to be present. When gauged according to the
\ion{N}{5}/\cfour, \ion{N}{5}/\ion{Si}{4} ratios, \mnhya\ falls in the gap
between the extreme objects BY~Cam and \vori, and the remaining 9 mCVs
available to Bonnet-Bidaud \& Mouchet (1987), all of which have \ion{N}{5}
$\lambda1240$/\cfour\ $\lambda1549<1$, \ion{N}{5} $\lambda1240$/\ion{Si}{4}
$\lambda1397<3$.  To that latter group we can add UZ~For and ST~LMi (Paper
II), and \voph\ and \vaql\ (Szkody \& Silber 1996; Friedrich, Staubert, \& la
Dous 1996a; this study -- Table~3).  Thus, the emission-line spectra of
BY~Cam, \vori, and to a lesser extent, \mnhya, remain anomalous.

\subsection{\voph\ (1H 1752+081)}

With the lowest mean surface field yet measured for any AM Her system
($B\sim7$~MG), \voph\ is probably synchronized only because it is also very
compact ($P=1.88$~hr).  The suggestion that the binary may house a small
accretion disk (Silber et al. 1994) has not found support in subsequent
studies, all of which indicate a system accreting solely via a stream (Barwig,
Ritter, \& B\"arnbantner 1994; Ferrario et al. 1995; Hessman et al. 1997;
Simi\'c et al. 1998; Steiman-Cameron \& Imamura 1999).  With such a weak
field, circular polarization has eluded detection, and thus far there is no
indication of the orientation of the white dwarf within the binary.

The UV light curves shown in Figure~3 record one clean ingress but only the
latter portion of egress\footnote{Note that the ephemeris of Barwig et al.
(1994) places $\varphi=0.0$ prior to the center of minimum light, as has been
found in other studies.}. In the expanded plot shown as Figure~10, ingress in
the optical and NUV is seen to be composed of a well-resolved ledge beginning
at $\varphi=0.972$ and a final decline to minimum light at $\varphi=0.993$.
From nearly coincident optical and X-ray photometry, Steiman-Cameron \& Imamura
(1999) showed that the X-ray eclipse is complete, with ingress occurring
during the optical ledge and X-ray egress near $\varphi=0.03$.  The totality
of eclipse at short wavelengths is confirmed by our NUV, FUV, and \cfour\
light curves.  In addition, the dereddened ($A_V=0.25\pm0.05$~mag) eclipse
spectrum shown in Figure~11 exhibits neither significant continuum nor line
emission. The optical decline that extends from $\varphi\sim0.972$ to 0.993
implies a source much larger than the white dwarf, and was interpreted by
Steiman-Cameron \& Imamura (1999) to be the extinguishing of the prograde
portion of the accretion stream.  The idea is verified by the similarity to
our \cfour\ light curve over this phase interval.  We return to this point
below.

In Figure~10 we identify contact points $c_1$ and $c_2$ with ingress and egress
of the ledge in the NUV, and measure a time interval $19\pm4$~s or
$\Delta\varphi=0.0028\pm0.0006$.  This is significantly shorter than the
$\sim$50~s duration measured in a broad optical bandpass by Steiman-Cameron \&
Imamura (1999), but is still comparable in size to the disk of a white dwarf
($\sim$$9\times10^8$~cm), and we follow their assignment as such. Within this
interval, points $c_1\arcmin$ and $c_2\arcmin$ mark the contacts of a
precipitous drop in the FUV flux which is strong evidence for an
accretion-heated spot on the stellar surface. The ingress interval of
$3.0\pm1.5$~s implies a linear distance of $\sim$$1.4\times10^8$~cm, or a
covering factor for a circular spot of $f_{\rm spot}\sim0.008$.  The amplitude
of the drop, $\sim$4.4 counts sample$^{-1}$, corresponds to a decline in
(dereddened) flux of $\Delta F_\lambda=7.7\times10^{-15}$ ergs cm$^{-2}$
s$^{-1}$ \AA$^{-1}$.  For $D=150\pm27$~pc (Silber et al. 1994), the spot
brightness temperature is $T_{\rm spot}\sim90,000$~K.

Out of eclipse, the {\it HST\/} light curves covering $0.9<\varphi<0.15$ show
a nearly constant flux level in the NUV and white-light bands, a slow decline
in the FUV, and steady rise in line emission.  These features can be explained
by a geometry in which the dominant pole is receding toward the stellar limb
during the eclipse interval, causing a reduction in FUV flux due to
foreshortening, but a rise in the line emission as the roughly radial,
optically-thick funnel presents an increasing cross-section to our view.  The
implication is that the orientation of the magnetic axis of the white dwarf
within the binary is more closely orthogonal to than it is aligned with the
stellar line of centers.

The protracted ingress of the accretion stream, which extends nearly to phase
zero or some 140~s beyond eclipse of the white dwarf, is a valuable diagnostic
of the extent that the accretion stream overshoots in azimuth the white
dwarf.  Such a path is of course expected for gas traveling in a purely
ballistic trajectory (Lubow \& Shu 1975). The simultaneous and rather abrupt
fall to zero light in all bands indicates that the final ingress truly defines
an ``edge'' to the gas, as opposed to a gradual thinning out (cf. the eclipse
profiles of \vori\ in Figures 1 and 6). The implied extension of the emitting
gas above the white dwarf is $7\times10^9$~cm, or $\sim$9~$R_{\rm wd}$. Lubow
\& Shu characterize stream paths for gas unaffected by magnetic fields
according to the point of closest approach from the accreting star,
$\omega_{min}$, and the radius of the disk that would ensue, $\omega_d$.  For
a non-disk system, we would expect that the maximum projected altitude of the
stream above the white dwarf at the time of eclipse would lie intermediate
between these characteristic heights. Using parameters estimated for \voph\
($M_1=0.9~M_\sun$, $M_2=0.185~M_\sun$; Barwig et al. 1994), we find
$\omega_{min}\approx6\times10^9$~cm and $\omega_d\approx9\times10^9$~cm, which
indeed bracket the measured extent above. It should be noted that because of
the dynamical nature of the stream/magnetosphere interaction, this height
might change with time, e.g. with varying accretion rate (cf. the variable,
$4-5~R_{\rm wd}$ height measured by Steiman-Cameron \& Imamura 1999).
Variations in the shape of stream egress in the interval $0.03<\varphi<0.05$
can be appreciated by an intercomparison of the curves of Steiman-Cameron \&
Imamura (1999) and to the white-curve in Figure~3.

\subsection{\vaql\ (RX J1940.1$-$1025)}

\subsubsection{Background}

Asynchronous AM Her systems have proven to be interesting puzzles.  The
prototype, V1500 Cyg (Stockman, Schmidt, \& Lamb 1988), was unusually
straightforward because it had already accumulated a rich photometric record
due to its 1975 nova outburst (e.g., Patterson 1979) and the system accretes
continuously onto both magnetic poles.  Among the subsequent 3
discoveries\footnote{BY Cam; V1432 Aql; CD
Ind=RX\,J2115$-$5840=EUVE\,J2115$-$586}, it has been more common to find
periods that are sidebands and/or harmonics of the underlying $P_{\rm spin}$
and $P_{\rm orb}$ due to pole-switching and migration of the accretion
footpoints through the lap period (see Ferrario \& Wickramasinghe 1999 and
Mouchet, Bonnet-Bidaud, \& de Martino 1999 for recent discussions).  The
bright and highly-polarized BY~Cam, for example, required study over a period
of several years, including an intense photometric campaign, to clearly
isolate the spin period of the white dwarf (Silber et al. 1992, 1997). The
presence of an eclipsing system among the known asynchronous mCVs is a
particular bonus because successive light curves portray an advancing
configuration between the stream and stellar magnetosphere.  An extensive
observational program could in principle track not only the accretion
footpoint(s) across the white dwarf but also the gas trajectory which gives
rise to those variations.  Such information would be invaluable in forming a
general understanding of the processes which couple the accreting gas to the
magnetic field.  \vaql\ thus occupies a position of unique importance for all
mCVs.

Because it is eclipsing, the binary period of \vaql\ is unequivocal at
3.365~hr.  However, controversy immediately arises over the origin of the
eclipses, and they have been more commonly labeled ``dips'' to reflect this
uncertainty.  Watson et al. (1995) noted phase jitter as well as an offset
between the optical/X-ray dips and the velocity zero-crossing of the emission
lines in proposing that the dips are due to obscuration by a high-density
accretion stream. An extensive photometric campaign by Patterson et al. (1995)
led to the identification of a superposed periodicity at 3.375~hr, taken to be
the spin period of the white dwarf and implying a lap period of 50~d. However,
the secondary star was preferred as the eclipsing object.  Several ensuing
studies (e.g., Friedrich et al. 1996b; Staubert et al. 1995; Geckeler \&
Staubert 1997) have supported and extended the duality of periods, but have
returned to an eclipse by the stream (but see Geckeler \& Staubert 1999).
Recently, even the period identification was called into question after
another variation at $\sim$$1\over3$ the orbital period was noted in archival
X-ray data (Mukai 1998).  The defining measurement -- detection of
phase-modulated circular polarization -- consists only of unpublished data
(from Buckley, quoted by Watson 1995) which appears to support the period
assignment of Patterson et al. (1995).

\subsubsection{New Results}

Four epochs of FOS spectroscopy were attempted of \vaql; unfortunately the
final observation was lost due to an on-board tape recorder failure.  Overall,
the data span $\sim$$1\over3$ of a 50~d lap cycle. As shown in Figure~4, dip
ingress and egress were each recorded on only one occasion. The light curves
vary enormously among themselves and are quite unlike those of the other
eclipsing systems.  At all epochs a downward inflection is apparent near
$\varphi=0.96$, at approximately the phase of optical ingress (e.g., Watson et
al. 1995).  However, the duration of this transition varies with epoch and FOS
waveband from $\Delta\varphi\sim0.03$ (360~s) for all continuum bands on 19
Aug. 1996 to $\Delta\varphi\sim0.005$ (60~s) for the FUV approximately 2 weeks
later.  The dips approach totality only in the FUV, and even here significant
signal remains at minimum light for all but the 3 Sep. observation.  The phase
of optical egress, $\varphi=0.035$, is barely covered on 29 Aug., but there is
no indication of a sharp inflection in any of the spacecraft bands at this
time; indeed the data are consistent with a smooth rise in count rate from the
brief minimum around $\varphi=0.98$. By contrast, the emission-line channel
shows no evidence of dips at all, with only a gradual, ill-defined fading
through the observation of 19 Aug.  The data from this epoch were also
searched for spectral shifts in the centroid and shape of \cfour\
$\lambda$1549 during the ``ingress'' period $0.96\lesssim\varphi\lesssim0.00$,
such as might occur from the eclipse of a disk.  No such variations were
detected to a limit of about one diode ($\sim$600~km~s$^{-1}$).

The top panel of Figure~12 presents the pipeline-processed spectrum of \vaql\
averaged over all epochs.  The plot basically resembles the single {\it IUE\/}
spectrum obtained by Friedrich et al. (1996a).  However, two new features are
evident in the much higher quality FOS data: a prominent 2200~\AA\ depression,
and the presence of \lalpha\ in absorption.  The interstellar absorption
feature can be erased cleanly for a visual extinction $A_V=0.7\pm0.1$~mag.
Eclipse ($\varphi=0.974-0.006$) and post-eclipse ($0.027-0.103$) spectra
corrected for this effect are shown in the bottom panel of Figure~12, both
taken from the run on 29 Aug. 1996, and line strengths computed from the mean
dereddened spectrum are listed in Table~3. A total extinction of $A_V=0.7$~mag
is by far the largest found among the objects studied here, but in good
agreement with a value of 0.6~mag derived from the Burstein \& Heiles (1982)
maps for $l=29\fdg0$, $b=-15\fdg5$.  The implied neutral hydrogen column is
$N_H\sim1.3\times10^{21}$~cm$^{-2}$, and the corresponding equivalent width
(EW) of damped \lalpha\ is $\sim$25~\AA\ (Wolfe et al. 1986).  Thus, the
interstellar medium is an important contributor to the deep absorption feature
we detect.  We have added a sequence of panels to the light curves in Figure~4
depicting the EW of the \lalpha\ line. In these plots, EW $>0$ corresponds to
an absorption line, EW $<0$ indicates emission, and the measurements have been
smoothed with a running-mean of width 40~s ($\Delta\varphi=0.003$).  This
index measures the net strength of the feature, since these low resolution
data do not resolve the absorption component from any emission line due to the
gas stream.  Note that the feature shows roughly constant EW except for phase
intervals when the FUV flux is near zero -- i.e.
$0.965\lesssim\varphi\lesssim0.000$ during the deep dips of 29 Aug. and 3 Sep.

\subsubsection{The UV Continuum Source}

The dereddened out-of-eclipse spectrum in Figure~12 exhibits a continuum that
rises smoothly to the blue, somewhat flatter than Rayleigh-Jeans.  When
compared to blackbodies, a temperature of $T\sim35,000$~K is found to
reproduce the UV slope and not exceed the dereddened optical continuum of the
only published spectrum of \vaql\ that is provided in absolute units
(Patterson et al. 1995).  To compute the size of the object with this effective
temperature, the distance estimate of $\sim$230~pc (Watson et al. 1995) must
be reduced by 16\% to $\sim$200~pc to account for the implied 0.3~mag of
$I$-band extinction, where detection of the M4 V secondary was made.  For a
circular object, the UV-optical source is then estimated to have a radius
$R\sim1.3\times10^9$~cm. If white dwarf model atmospheres are used instead
(Bergeron 1994), a somewhat lower temperature, $T=25,000-30,000$~K, and larger
size, $R\sim1.6\times10^9$~cm, ensue. The model atmospheres also predict a
\lalpha\ absorption feature with an EW of 30~\AA\ or more which would blend
with the interstellar component at this resolution.

The similarity of the derived source size to the disk of a white dwarf is very
attractive, and at face value suggests a low-mass star ($M_{\rm
wd}=0.15-0.25~M_\sun$).  However, there is significant latitude in the combined
uncertainties of the extinction correction and distance estimate.  Moreover,
the white dwarf need not supply the entire observed continuum in either the UV
or optical (note the residual flux level in the eclipse spectrum of
Figure~12).  Thus, a smaller stellar size (larger mass) is possible. A
somewhat smaller size, $R\sim6\times10^8$~cm, is also indicated by the 40~s
duration of the most abrupt FUV eclipse ingress on 3 Sep. 1996 (Figure~4). The
corresponding luminosity for emission by an entire white dwarf with $T_{\rm
wd}=30,000-35,000$~K is $L_{\rm wd}\sim0.2-0.5~L_\sun$, of which $\sim$2\% is
intercepted by the secondary star. This illumination amounts to at most 50\%
of the $0.02~L_\sun$ nuclear luminosity of an M4 dwarf (cf. V1500 Cyg --
Schmidt \& Stockman 1995), so the brightness modulation at the orbital period
due to reprocessing would be expected to be modest\footnote{And dependent on
the unknown system inclination.}. Indeed, the average light curve of Patterson
et al. (1995) shows a $\sim$0.5~mag variation away from the dip feature,
peaking near $\varphi=0.5$, as is appropriate for an origin on the inner
hemisphere of the companion (see the discussion of this point by Patterson et
al.). A contribution from the white dwarf and illuminated secondary is also
consistent with the weak level of optical circular polarization measured for
\vaql\ ($|v|\sim2$\%, Friedrich et al. 1996b; Watson 1995).  Because a hot
white dwarf can account for the basic characteristics of the UV-optical
spectral energy distribution and is consistent with the mean optical light
curve and polarization, we adopt this explanation throughout the remainder of
the paper.

\subsubsection{Origin of the Dips}

With the observed energy distribution from the UV through the optical
dominated by the hot white dwarf, we now investigate the origin of the
photometric dips. Discussion on this topic has centered around the regularity
of the features, their depth in various wavelength regions, the duration of
ingress/egress, and their phasing relative to other indicators of orbital
motion. Secondary star proponents point out the very predictable dip ephemeris
and phasing relative to the emission-line radial velocities (Patterson et al.
1995).  Gas stream advocates, on the other hand, note an energy dependence of
eclipse depth in X-rays and the ingress/egress intervals, which exceed those
expected for the sharp-edged eclipse of a white dwarf (Watson et al. 1995;
Friedrich et al. 1996b; Geckeler \& Staubert 1997).

New information on this question is provided by the FOS spectroscopy.  First,
the dips in Figure~4 are seen to be highly variable in both {\it shape\/} and
{\it depth\/} from epoch to epoch.  Note in particular the absence of a
defined ingress on 19 Aug., the lack of an expected egress on 30 Aug., and the
variable depth at minimum light.  Such behavior is natural for absorption due
to a dense, ill-defined gas stream but at odds with a stellar eclipse. This is
particularly true for the FUV channel, which from our prior conclusions is
dominated by the hot white dwarf.  Indeed, based on our results for UZ For
(Paper II) and the other systems studied here, it is difficult to imagine any
source of appreciable light in the $\sim$$1250-1520$~\AA\ region that would
not be located on or very near the white dwarf.  Second, the UV emission lines
are essentially {\it undiminished\/} through the absorption dips.  This fact
has already been noted for the optical lines by Watson et al. (1995) and
contested by Patterson et al. (1995; see their Figures~12 and 13). The issue
revolves around the various line components and their places of origin.  In
the case of the high-ionization lines of \cfour\ and \ion{He}{2}, the lack
of eclipse features in Figure~4 is unequivocal.  We also point out that if the
white dwarf were being eclipsed by a secondary which fills its Roche lobe, the
observed duration of the dip (820~s or $\Delta\varphi=\pm0.035$) requires an
inclination $i\ge78\arcdeg$.  In this case, the eclipse track crosses behind
the secondary disk $\sim$$2\over3$ of the way from the center.  The remainder
of the disk would be expected to obscure line-emitting gas in the lower
accretion streams(s) out to $\sim$7~$R_{\rm wd}$ above the orbital plane for a
period of up to $\sim$800~s.  No such reduction in emission-line light is
evident.

We therefore conclude that the absorption dips in \vaql\ are due to the
intervention of a dense portion of the accretion flow. In the context of
emission by a white dwarf disk and eclipse by a stream, the curious ``choppy''
appearance to the light curves {\it at all orbital phases\/} (Watson et al.
1995; Patterson et al. 1995) is suggestive of persistent obscuration by gas
clumps, such as might occur in a circulating accretion ``ring''.  A similar
structure is inferred to exist in the asynchronous mCV V1500 Cyg from its
ever-present sinusoidal polarization curve that implies 2-pole accretion
(Schmidt, Liebert, \& Stockman 1995).  We also note that the lack of an
identifiable eclipse by the secondary implies that the inclination of \vaql\
is $i<73\arcdeg$.  A very rough estimate can also be gained from the observed
radial velocity half-amplitude of the narrow emission lines (160~km~s$^{-1}$;
Patterson et al. 1995), which for reasonable white dwarf and secondary star
masses ($\sim$0.6 and 0.3~$M_\sun$, respectively) suggests $i\sim40\arcdeg$.

\section{\vaql\ as a Recent Nova}

The link between the nova event and asynchronism was established with the
discovery that the 1975 nova system V1500 Cyg is an mCV (Stockman et al. 1988).
In support of this position, Patterson et al. (1995) has pointed out the
vanishing likelihood that all 4 asynchronous systems would be synchronizing for
the first time, given that the spin-orbit period difference is universally
$\sim$1\% and that, when measured, locking timescales are all in the range
$10^2-10^3$~yr (Schmidt et al. 1995; Mason et al. 1995; Geckeler \& Staubert
1997).  The likely alternative is that most, if not all, of the asynchronous
AM Her systems were previously phase-locked and that the magnetic connections
were broken in recent nova events.

Only in the case of V1500 Cyg is the post-nova classification certain. The
object has an extremely rich observational history post-1975, the white dwarf
temperature has been measured at $T_{\rm wd}\sim90,000$~K and decreasing, and
the object lies at the center of an expanding debris shell.  None of the other
asynchronous systems have been associated with a nebula, despite searches. The
likelihood that the aforementioned abundance anomalies of BY~Cam (\S3.2.) are
indicative of mixing with nova-processed material is reduced now that the
synchronized systems \vori\ and \mnhya\ have been found to show elevated
\ion{N}{5} line strengths.  A nova event in the recent past of the 8~hr-period
\vori\ appears particularly improbable, as it should be especially easy to
disrupt and slow to resynchronize.

Excluding V1500 Cyg, the temperature of $\sim$35,000~K that we have measured
for the white dwarf in \vaql\ is far and away the hottest yet found on an AM
Her system (Sion 1999; G\"ansicke 1999).  When all CVs are considered, white
dwarfs this hot have been detected {\it only\/} in dwarf nova and nova-like
classes (Sion 1999), both of which are thought to be eruptive.  It is
therefore tempting to classify \vaql\ as the second post-nova AM Her system.
Using the cooling calculations of Prialnik (1986) as a guide, the current
stellar luminosity of $0.2-0.5~L_\sun$ for \vaql\ vs. $\sim$5~$L_\sun$ for
V1500 Cyg (Schmidt et al. 1995) suggests that the interval since the last
eruption of \vaql\ is in the range $\Delta t=75-150$~yr.  At a distance
$D\sim200$~pc and an expansion velocity $v=1000$~km~s$^{-1}$, the shell could
now be several arcminutes across. Because the surface brightness of a nebular
remnant scales as $(v\Delta t)^{-5}$ (Cohen 1988)\footnote{The dependence is
even steeper if the nebula is ionization-bounded.}, direct imaging of it might
be very difficult (see also Wade 1990).  We note that the faintness and lack
of a hot blackbody component in the near-UV spectrum of BY~Cam (Bonnet-Bidaud
\& Mouchet 1987; Shrader et al. 1988) suggest that, if it is also a post-nova
system, it is considerably older still and thus detection of a shell would be
hopeless.

The fly in the ointment for a post-nova explanation of \vaql\ is the fact that
in this case $P_{\rm spin}>P_{\rm orb}$. The mechanism outlined by Stockman et
al. (1988) for decoupling V1500 Cyg involved the transfer of a small amount of
orbital angular momentum to the bloated white dwarf during the common-envelope
phase of the nova.  When the photosphere contracted back within the binary and
eventually to a normal stellar radius, the two motions could be distinguished
observationally, and $P_{\rm spin}<P_{\rm orb}$. To achieve the \vaql\
configuration from an outburst in an initially synchronized binary requires
either decreasing the orbital period of the binary (such that the secondary
spirals inward) or increasing the spin period of the primary.  Shara et al.
(1986) has shown that the net effects of frictional angular momentum loss on a
secondary orbiting within the envelope results in a decrease in the binary
dimension $a$ only for slow novae with very low-mass secondaries ($M_2/M_{\rm
wd}\lesssim0.01$). Of course, for \vaql, it is not sufficient merely that
$\Delta a<0$, but that the resulting decrease in $P_{\rm orb}$ exceed the
decrease in $P_{\rm spin}$ due to spin-up of the envelope (the V1500 Cyg
effect). Thus, an explanation of the sense of asynchronism of \vaql\ purely in
terms of dynamical effects is not likely.

More promising is rotational braking of the white dwarf during the mass-loss
phase via the strong magnetic field (the magnetic propeller effect).  This
problem has been presented in many ways for single stars as well as members of
binary systems.  We choose the convenient formulation of Kawaler (1988), which
can be expressed in terms of the fractional change in rotational angular
momentum
\begin{equation}
{\Delta J \over J} = {5\over3}{\Delta M \over M_{\rm wd}} \Bigl({r_A \over R_{\rm wd}}\Bigr)^n_{\rm radial}~.
\end{equation}
Here, $\Delta M$ is the total mass lost in the eruption, $r_A$ (radial)
is the Alfv\`en radius out to which corotation is maintained for a radial
magnetic field line distribution, $n$ is an index that denotes the departure
from a radial field geometry, and we have approximated the moment of inertia
of the white dwarf by that of a uniform-density sphere.  A value $n=2$ denotes
radial field lines, while $n=3/7$ is appropriate for a dipole.  The expression
for $r_A$ provided by Kawaler (1988) can be rewritten using parameters
appropriate for a 0.6~$M_\sun$ white dwarf as
\begin{equation}
\Bigl({r_A \over R_{\rm wd}}\Bigr)_{\rm radial} \approx 1.2\times10^6~B_0^{4/3}~\dot M^{-2/3}
\end{equation}
with $B_0$ being the surface magnetic field in gauss and $\dot M$ in
gm~s$^{-1}$.  Assuming $B_0=3\times10^7$~G and continuous mass loss totalling
$\Delta M=10^{-4}~M_\sun$ over a wind phase of $\Delta t=10^7$~s (MacDonald,
Fujimoto, \& Truran 1985), we find a decoupling radius for radial field lines
$r_A~(\rm radial)\sim 15~R_{\rm wd}$, well inside the orbit of the secondary.
The total angular momentum lost is then ${\Delta J/J} = 0.001-0.06$, with the
two limits signifying dipolar and radial geometries, respectively.  The
consistency with the 0.3\% period difference currently observed for \vaql\ is
encouraging.

The above derivation is admittedly crude.  It serves, however, to illustrate
the possible importance of magnetic braking during the outburst.  In view of
this result, it might be better to view V1500 Cyg as an example, rather than a
template, of a nova eruption in an mCV, and given that the net outcome is the
result of a competition between at least two effects with magnitudes of a few
percent each, a system $P_{\rm spin}>P_{\rm orb}$ might also be expected. In
fact, with $r_A$ being most sensitive to the magnetic field strength and
duration of the wind phase ($\propto B_0^{4/3}\Delta t^{2/3}$), a very fast
nova like V1500 Cyg could experience spin-up, while a higher-field, slow nova
might be spun down.

\section{Conclusions}

Time-resolved {\it HST\/} UV spectroscopy of four eclipsing mCVs has enabled
us to identify:

\begin{itemize}

\item{Accretion-heated photospheric spots in \vori\ and \voph\ with covering
factors $f<0.015$ and $\sim$0.008, and blackbody temperatures $T_{\rm spot}>150,000$~K
and $\sim$90,000~K, respectively.}

\item{A cyclotron shock on \mnhya\ with $f\lesssim0.004$.}

\item{Prominent emission from the accretion stream in \vori\ and \voph.  A
protracted emission-line ingress in the latter object indicates the
stream penetrates the magnetosphere to a height of $\sim$9~$R_{\rm wd}$ before it
is funneled onto the magnetic poles.}

\item{Unusually strong \ion{N}{5} $\lambda$1240 in \vori\ and, to a lesser extent, \mnhya,
that is suggestive of an overabundance of nitrogen in the gas stream.}

\item{A spectral energy distribution for \vaql\ that strongly suggests
the dominant source of energy is a
hot ($T_{\rm wd}\sim35,000$~K) white dwarf.  The undiminished line
emission through eclipse indicates that the eclipse is caused by a
dense portion of the accretion stream, not the secondary star. The very hot white dwarf,
combined with the current asynchronous nature
and rapid timescale for relocking, suggests that \vaql\ underwent a nova
eruption in the past $\sim$$75-150$~yr.  A reversed sense of asynchronism,
with the primary star currently spinning up toward synchronism, is not
at odds with this history, if the white dwarf rotation
can magnetically couple to the ejecta during the dense wind phase of the
eruption.}

\end{itemize}

Time-tagged UV spectrophotometry is an important tool for understanding the
sizes and relative contributions of the accretion stream, hot spot, and white
dwarf photospheres in mCVs.  Both the Space Telescope Imaging Spectrograph
(STIS) and the Cosmic Origins Spectrograph (COS) offer sensitivities an order
of magnitude greater than the FOS.  Unfortunately, with the limited aperture
of {\it HST\/}, it will be difficult to resolve objects smaller than 10\% of
the white dwarf radius.  Future UV telescopes, with apertures exceeding 4~m
and high efficiency, energy-resolving detectors will be ideal facilities for
discerning details of structures as small as the accretion shock itself.

\acknowledgments

Special thanks go to Carolyn Pointek for assistance with the initial data
analysis and to R. Wagner for spearheading the development of high-speed CCD
photometry at the MMT and for lending assistance at the telescope. P. Bergeron
kindly provided DA model atmosphere spectra. Support was provided through NASA
grant GO-6498 from the Space Telescope Science Institute, which is operated by
the Association of Universities for Research in Astronomy, Inc., under NASA
contract NAS 5-26555.  Ground-based studies of magnetic stars and stellar
systems are funded by NSF grant AST 97-30792 to GDS.

\clearpage

\begin{deluxetable}{lccccccl}
\tablenum{1}
\tablewidth{0pt}
\tablecaption{System Parameters}
\tablehead{\colhead{System} &
\colhead{$P_{\rm orb}$ (hr)} &
\colhead{$D$ (pc)} &
\colhead{$A_V$~(mag)} &
\colhead{$\beta$\tablenotemark{a}} &
\colhead{$\psi$\tablenotemark{b}} &
\colhead{$B$ (MG)} &
\colhead{Refs.} }
\startdata
\vori       & 7.98271   & $>$500($>$1500?) & $0.15\pm0.05$ & $\cdots$              & $\sim$90\arcdeg & 61       & 3,5,7,10 \\ %
\mnhya      & 3.38985   & $300-700$        & $0.10\pm0.05$ & $\sim$30\arcdeg       & $\sim$0\arcdeg  & 20: 42:  & 2,9,13 \\
\voph       & 1.88280   & $150\pm27$~\,    & $0.25\pm0.05$ & $15\arcdeg-30\arcdeg$ & $\cdots$        & 7        & 1,4,11 \\
\vaql       & 3.36564   & $180-250$        & $0.7\pm0.1$   & ~\,$\sim$58\arcdeg?   & asynchronous    & $\cdots$ & 6,8,12,13 \\
\enddata
\tablenotetext{a}{Colatitude of accretion region}
\tablenotetext{b}{Azimuth of accretion region}
\tablerefs{
(1)~Barwig et al. (1994);
(2)~Buckley et al. (1998b);
(3)~Buckley \& Shafter (1995);
(4)~Ferrario et al. (1995)
(5)~Garnavich et al. (1994);
(6)~Geckeler \& Staubert (1997);
(7)~Harrop-Allin et al. (1997)
(8)~Patterson et al. (1995);
(9)~Ramsay \& Wheatley (1998);
(10)~Shafter et al. (1995);
(11)~Silber et al. (1994);
(12)~Watson et al. (1995);
(13)~this paper.
}
\end{deluxetable}

\begin{deluxetable}{llccl}
\tablenum{2}
\tablewidth{0pt}
\tablecaption{Observing Log}
\tablehead{\colhead{System} & \colhead{UT Date} & \colhead{HJD Start$-$End} & \colhead{Type\tablenotemark{a}} & \colhead{Comment} \\
& & \colhead{(2450000+)} }
\startdata
\vori       & 1996 Aug 4  & $299.9656~\,-.9725$\,~ & OS & $V\sim16.5~@~\varphi=0.33$ \\
(RX J0515.6+0105) & 1996 Aug 12 & $307.80673-.81753$ & US & \\
        & 1996 Aug 22 & $318.12780-.13943$ & US & \\
        & 1996 Aug 22 & $318.14311-.15359$ & US & \\
        & 1996 Aug 24 & $320.13874-.15037$ & US & \\
        & 1996 Aug 24 & $320.15405-.16453$ & US & \\
        & 1996 Aug 26 & $322.14953-.16116$ & US & \\
        & 1996 Aug 26 & $322.16484-.17532$ & US & \\
        & 1996 Sep 16 & $343.39618-.40701$ & US & \\
        & 1996 Sep 27 & $353.72920-.73968$ & US & \\
        & 1996 Oct 1  & $357.73670-.74834$ & US & \\
        & 1996 Oct 1  & $357.75202-.76249$ & US & \\
        & 1996 Oct 25 & $382.49924-.51064$ & US & \\
        & 1996 Oct 26 & $382.51433-.52480$ & US & \\
        & 1996 Dec 3  & $420.8318~\,-.8422$\,~ & OS & $V\sim15.9~@~\varphi=0.73$ \\
\\
\mnhya      & 1996 Jan 27 & $109.822~~~-.970$~~~ & P,Ph & $v=-17\%$ to $+5\%$ \\
(RX J0929.1$-$2404) & 1996 Mar 10 & $152.790~~~-.799$~~~      & Ph & 128~ms time samples \\ % look this up!
        & 1996 Oct 14 & $371.28629-.29618$ & US & Extremely faint \\
        & 1996 Oct 14 & $371.29987-.30455$ & US & Extremely faint \\
        & 1996 Oct 15 & $372.02~~~~\,-.03$~~~~\, & OS & $V>20$ \\
        & 1996 Dec 4  & $421.98~~~~\,-.99$~~~~\, & OS & $V\sim16.9$ \\
        & 1996 Dec 24 & $442.05640-.06629$ & US & \\
        & 1996 Dec 24 & $442.06997-.07466$ & US & \\
\\
\voph       & 1996 Aug 3  & $299.08228-.09217$ & US & \\
(1H 1752+081)   & 1996 Aug 3  & $299.14747-.15217$ & US & \\
        & 1996 Aug 4  & $299.6543~\,-.6612$~\, & OS & $V\sim15.8~@~\varphi=0.42$ \\
        & 1996 Aug 30 & $325.74371-.75360$ & US & \\
        & 1996 Aug 30 & $325.75729-.76197$ & US & \\
\\
\vaql       & 1996 Aug 19 & $314.68951-.70027$ & US & \\
(RX J1940.1$-$1025) & 1996 Aug 29 & $325.34667-.35830$ & US & \\
        & 1996 Aug 29 & $325.36198-.37245$ & US & \\
        & 1996 Sep 3  & $330.25291-.26367$ & US & \\
        & 1996 Sep 9  & $335.86638-.87801$ & US & FGS/tape recorder failure \\
        & 1996 Sep 9  & $335.88169-.89217$ & US & FGS/tape recorder failure \\
\enddata
\tablenotetext{a}{US = UV spectroscopy; OS = optical spectroscopy;
P = optical polarimetry, Ph = optical photometry}
\end{deluxetable}

\begin{deluxetable}{lcccc}
\tablenum{3}
\tablewidth{0pt}
\tablecaption{UV Line Strengths\tablenotemark{a}~ ({\rm $\times10^{-13}$ ergs cm$^{-2}$ s$^{-1}$})}
\tablehead{\colhead{$A_V$~(mag)} & \colhead{$0.15\pm0.05$} & \colhead{$0.10\pm0.05$} & \colhead{$0.25\pm0.05$} & \colhead{$0.7\pm0.1$} }
\tablehead{\colhead{Line} & \colhead{\vori} & \colhead{\mnhya} & \colhead{\voph} & \colhead{\vaql}}
\startdata
\ion{C}{3}  $\lambda$1176 & $\cdots$~\, & $\cdots$~~ & 8.2~\,   & 25.8~\, \\
\lalpha\tablenotemark{b} \ $\lambda$1216 & 2.6~\,   & 3.9~\,   & 8.3~\, & Absorption \\
\ion{N}{5}  $\lambda$1240 & 17.4~~~  & 2.3~\,   & 13.3~~~   & 41.8~\,   \\
\ion{Si}{3} $\lambda$1298 & 1.2~\,   & 0.35     & 4.1~\,   & 4.5     \\
\ion{C}{2}  $\lambda$1335 & 0.32     & $\cdots$~\, & 3.8~\,   & 4.3  \\
\ion{O}{5}  $\lambda$1371 & 0.93     & \,0.14:  & $\cdots$~\, & 5.1  \\
\ion{Si}{4} $\lambda$1397 & 5.4~\,   & 1.3~\,   & 13.3~~~   & 23.8~\,   \\
\ion{C}{4}  $\lambda$1549 & 2.4~\,   & 0.76     & 29.4~~~  & 83.4~\,  \\
\ion{He}{2} $\lambda$1640 & 4.5~\,   & 0.94     & 7.2~\,   & 24.6~\,   \\
\ion{N}{4}  $\lambda$1718 & 1.6~\,   & \,0.09:  & $\cdots$~\, & 1.7  \\
\ion{Al}{3} $\lambda$1857 & 1.2~\,   & 0.25     & 3.8~\,   & 2.9     \\
\ion{C}{3}  $\lambda$2297 & \,0.29:  & $\cdots$~\, & 0.87     & 2.4   \\
\ion{He}{2} $\lambda$2383 & \,0.18:  & $\cdots$~\, & $\cdots$~\, & 1.1 \\
\enddata
\tablenotetext{a}{Extinction-corrected}
\tablenotetext{b}{May have a geocoronal emission component}
\end{deluxetable}

\clearpage

\begin{figure}%Figure 1
\includegraphics[bb=50 650 2 2, scale=.9, angle=0]{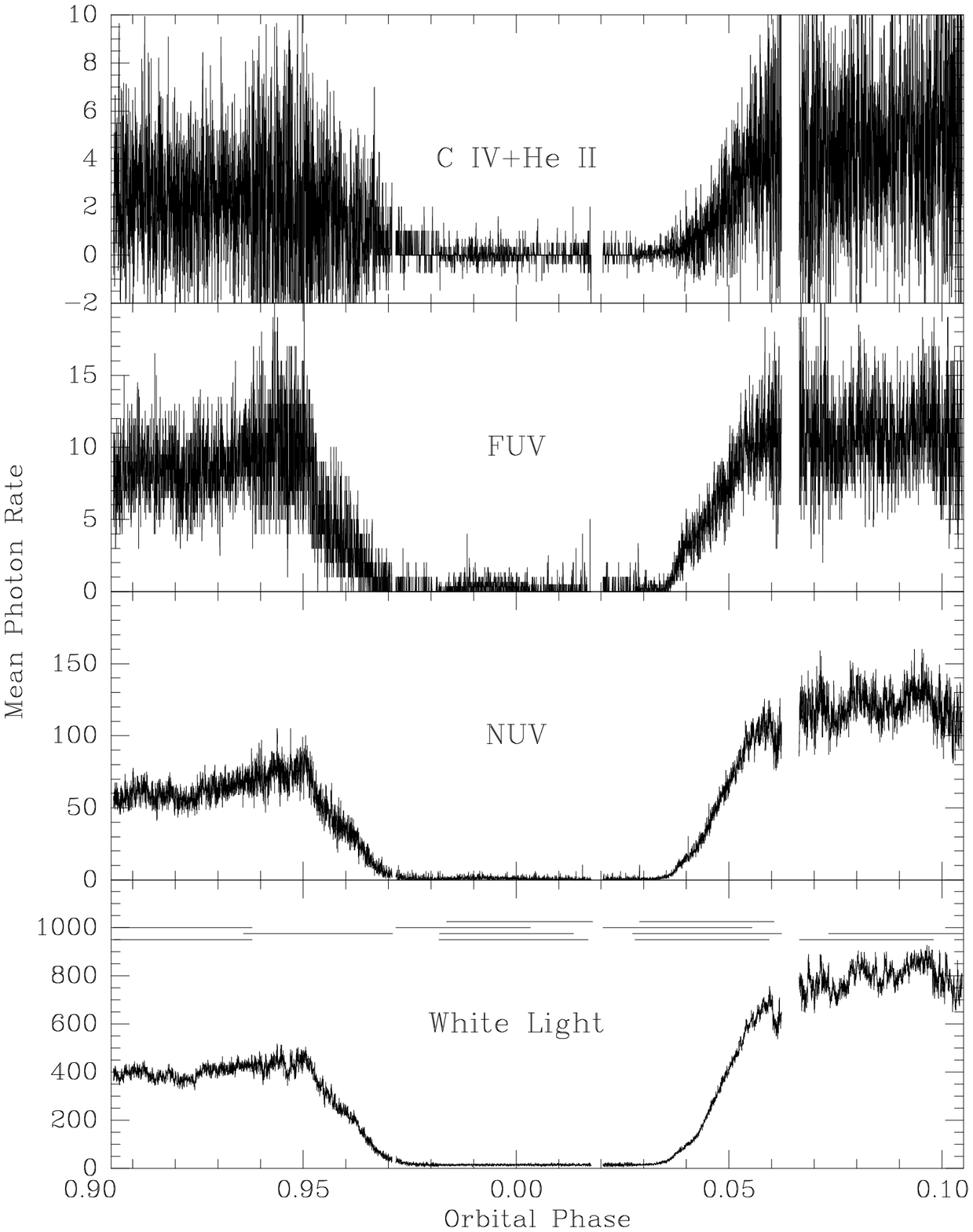}
\vskip7.truein
\caption{{\it HST\/} eclipse light curves averaged over all
epochs of the magnetic variable \vori. Panels represent: {\it
\cfour+\hetwo:\/} the \cfour\ $\lambda\lambda1548,1551$ plus \hetwo\
$\lambda1640$ emission lines; {\it FUV:\/} the $\lambda\lambda1255-1518$
continuum; {\it NUV:\/} the $\lambda\lambda 1945-2506$ continuum; {\it White
Light:\/} the zero-order image, a broad passband with $\lambda_{\rm
eff}\sim3400$\AA.  Horizontal line segments in the bottom panel indicate phase
intervals of the individual time series.  Rate units are detected photon
counts per 0.8201~s sample. }
\end{figure}

\clearpage

\begin{figure}%Figure 2
\includegraphics[bb=50 610 2 2, scale=.9, angle=0]{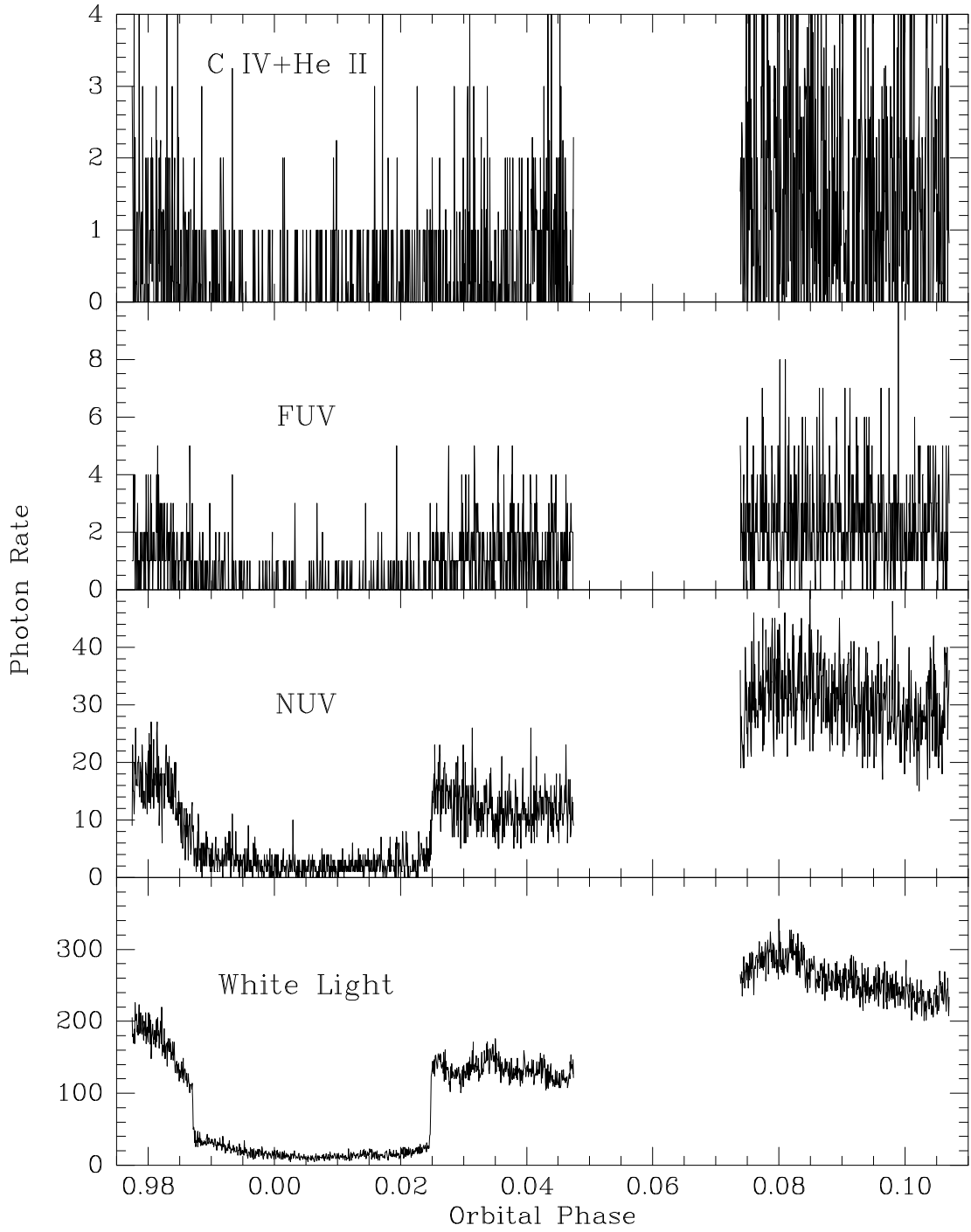}
\vskip6.5truein
\caption{As in Figure~1 for \mnhya.  Due to the extremely faint state encountered
in the initial visit, only data from the second pointing are shown here.}
\end{figure}

\clearpage

\begin{figure}%Figure 3
\includegraphics[bb=50 610 2 2, scale=.9, angle=0]{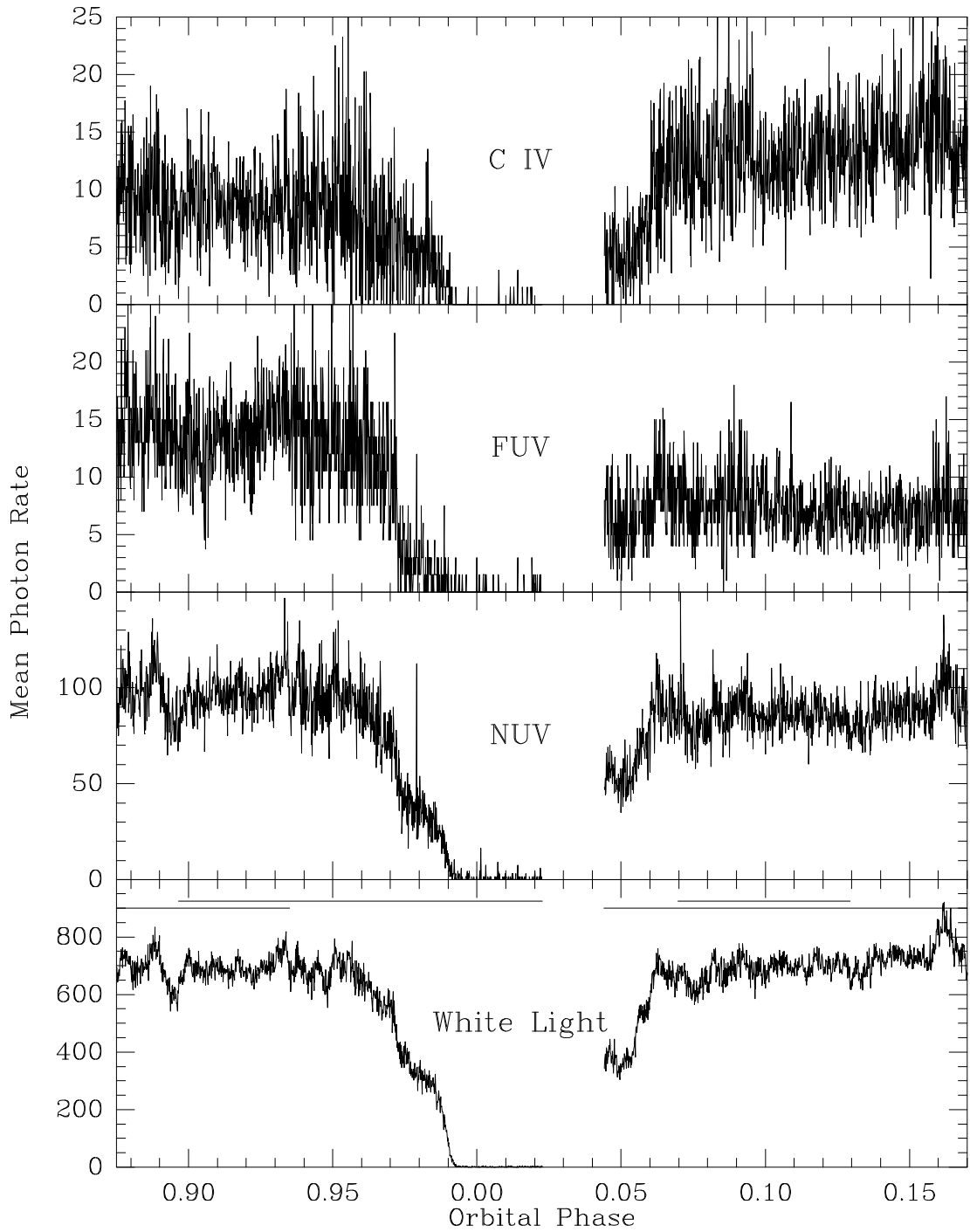}
\vskip6.5truein
\caption{As in Figure~1 for \voph.  With \cfour\
$\lambda$1549 dominating the UV emission-line spectrum, its flux alone is
displayed in the top panel.}
\end{figure}

\clearpage

\begin{figure}%Figure 4
\includegraphics[bb=0 670 2 2, scale=.9, angle=0]{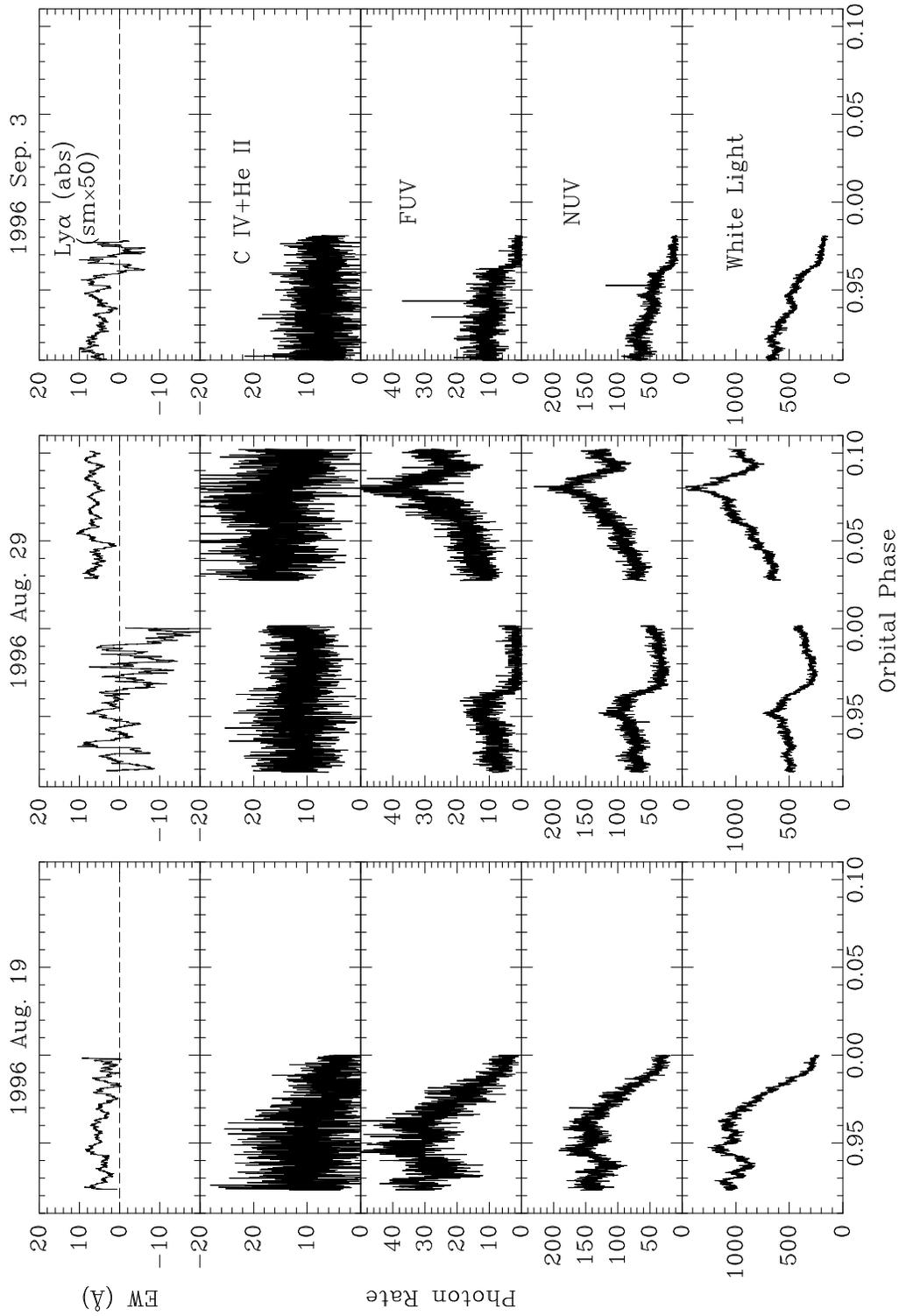}
\vskip7.7truein
\caption{{\it HST\/} eclipse light curves for 3 epochs of the
asynchronous system \vaql.   Included as a top series of panels is the
equivalent width of \lalpha, smoothed over 50 time samples (40~s
or $\Delta\varphi=0.003$).  EW $>0$ indicates absorption.}
\end{figure}

\clearpage

\begin{figure}%Figure 5
\includegraphics[bb=420 50 2 2, scale=.7, angle=-90]{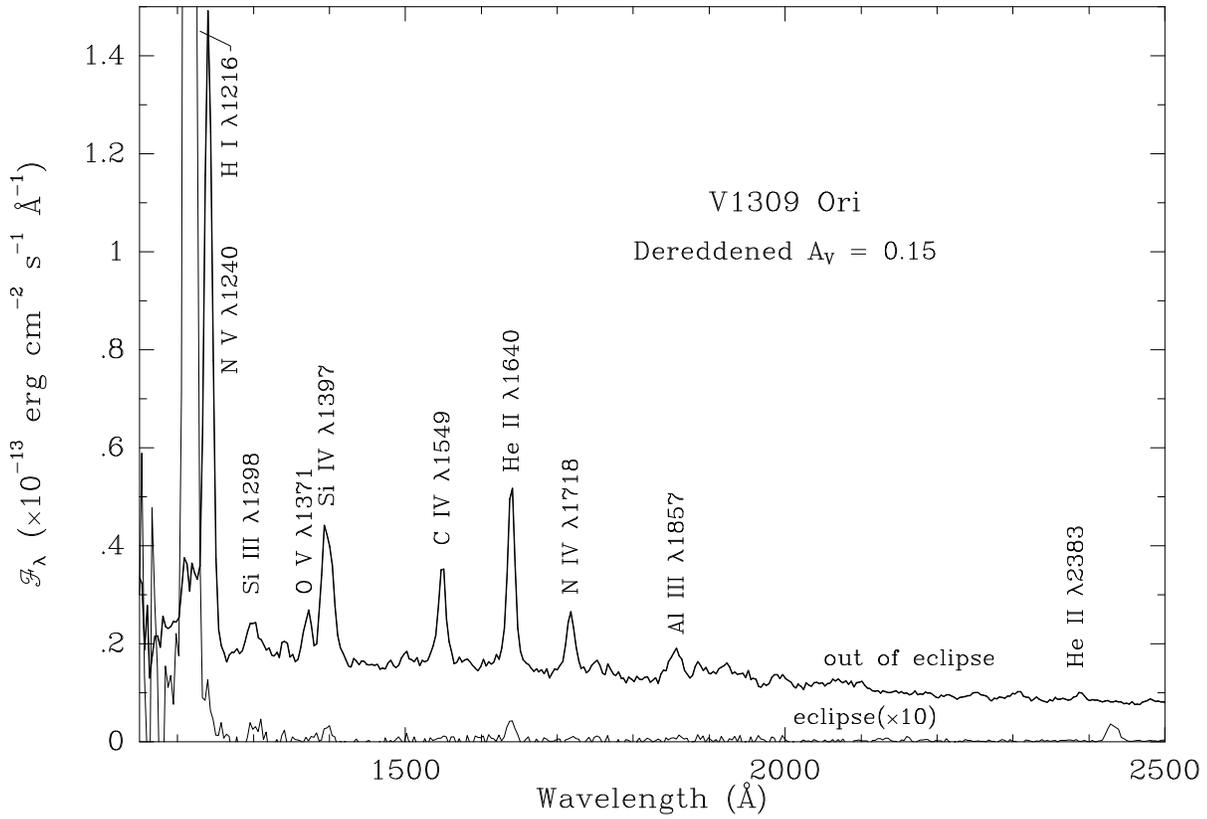}
\vskip1.truein
\caption{Mean in- and out-of-eclipse spectra for \vori,
corrected for interstellar extinction in the amount $A_V=0.15$~mag.  Note the
unusual strength of \ion{N}{5} $\lambda$1240 and \ion{N}{4} $\lambda$1718.}
\end{figure}

\clearpage

\begin{figure}%Figure 6
\includegraphics[bb=40 670 2 2, scale=.9, angle=0]{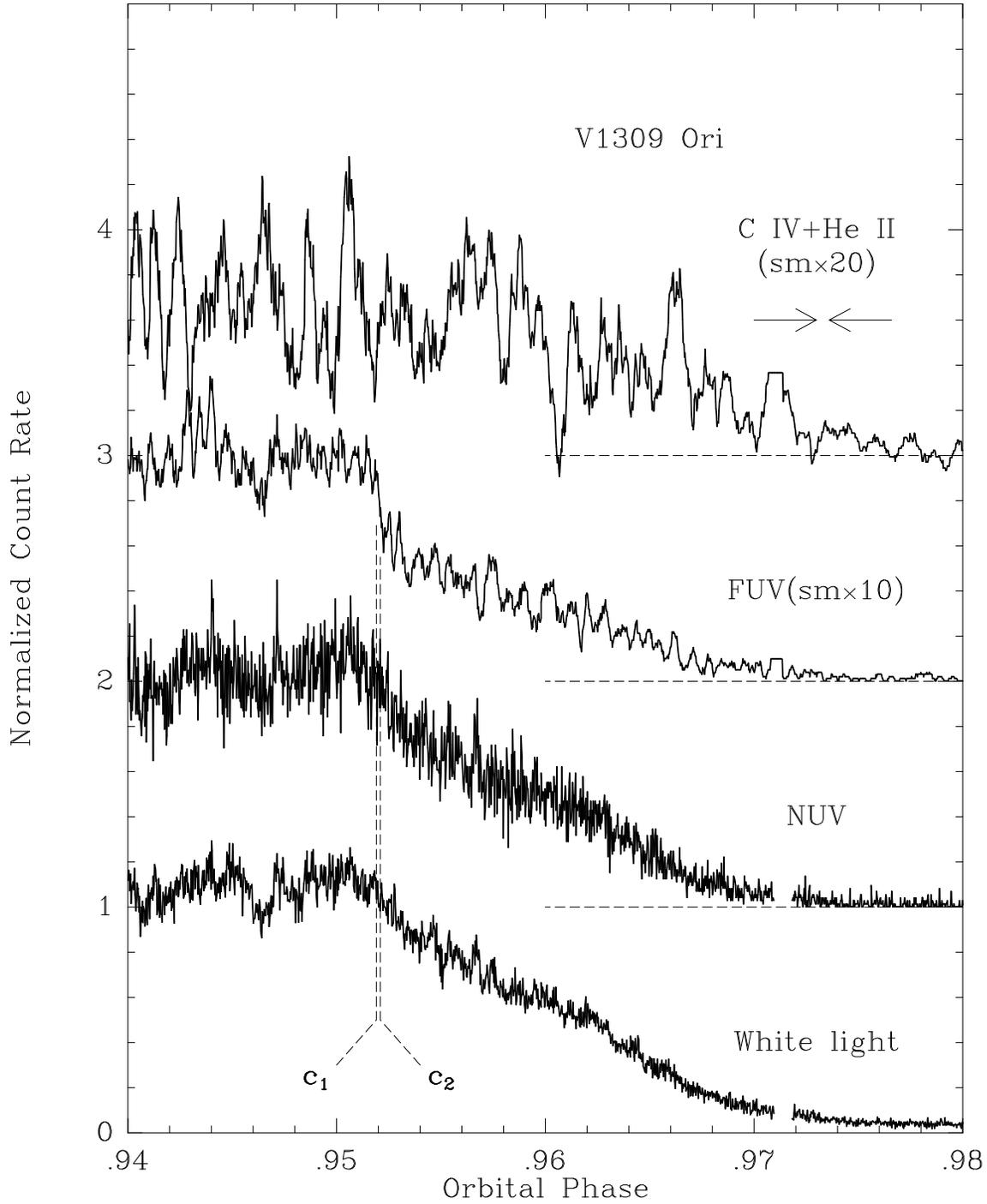}
\vskip7.3truein
\caption{Eclipse ingress for \vori.  The FUV channel has been
smoothed with a running mean of width 10 time steps (8.2~s) and the
emission-line band by a 20-step wide window (16.4~s or
$\Delta\varphi=0.0006$).  The width of the latter window is indicated by the
gap between opposing arrows. Eclipse contacts $c_1,~c_2$ in the FUV channel
indicate ingress of a hot accretion spot which is unresolved in the smoothed
data.}
\end{figure}

\clearpage

\begin{figure}%Figure 7
\includegraphics[bb=230 -20 2 2, scale=.6, angle=-90]{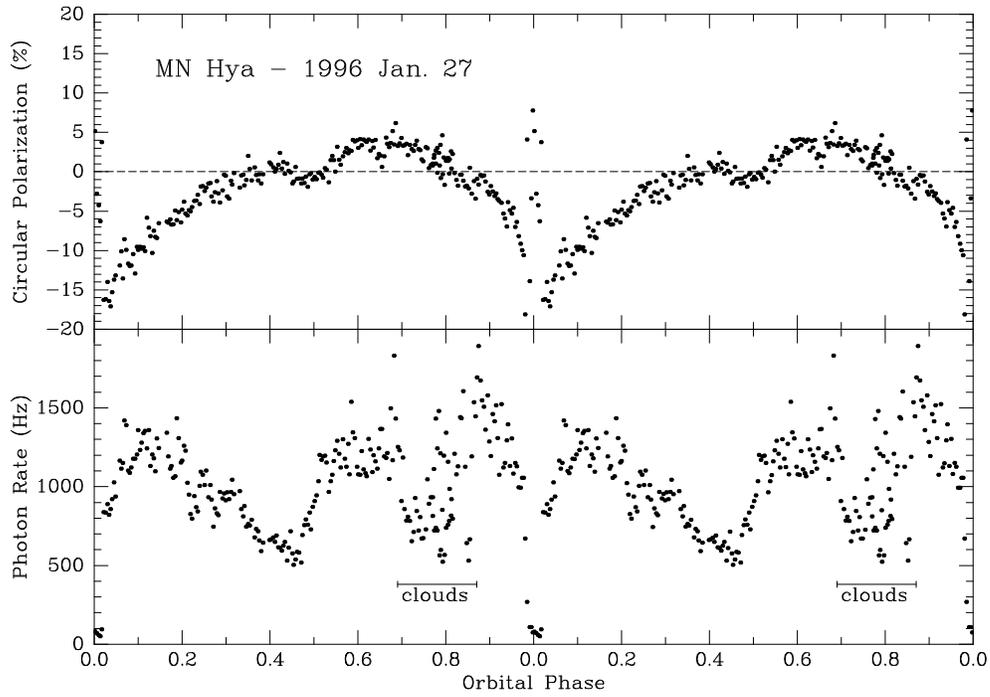}
\vskip2.3truein
\caption{Circular polarization and brightness in the optical
covering one orbital cycle of \mnhya.  Note the deep eclipse at $\varphi=0$
and the $\cal S$-wave in circular polarization around $\varphi=0.5$ associated
with the two poles rotating over the stellar limb. The data are plotted twice
for clarity.}
\end{figure}

\clearpage

\begin{figure}%Figure 8
\includegraphics[bb=350 -20 2 2, scale=.6, angle=-90]{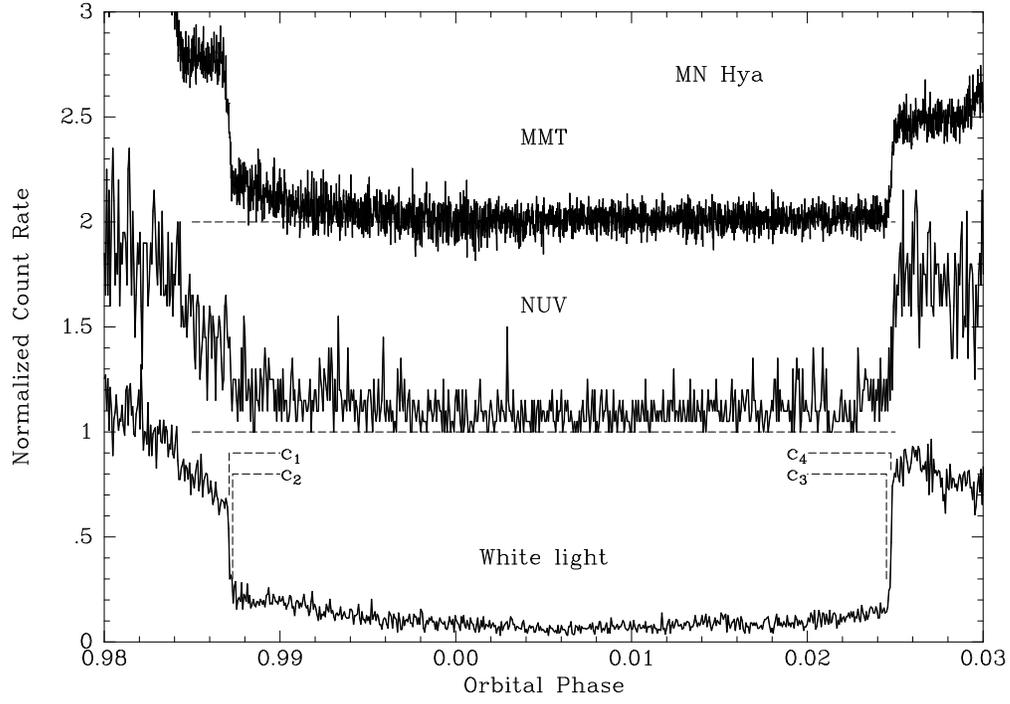}
\vskip1.4truein
\caption{Eclipse detail of \mnhya. The top light curve is an
optical light curve ($\lambda\lambda3500-6500$) taken on 1996 Mar. 10 at the
MMT with a CCD clocked in a programmed-readout mode.  Contacts mark the
ingress and egress of an optical-near UV-emitting spot assigned to cyclotron
emission from the accretion shock.}
\end{figure}

\clearpage

\begin{figure}%Figure 9
\includegraphics[bb=420 50 2 2, scale=.7, angle=-90]{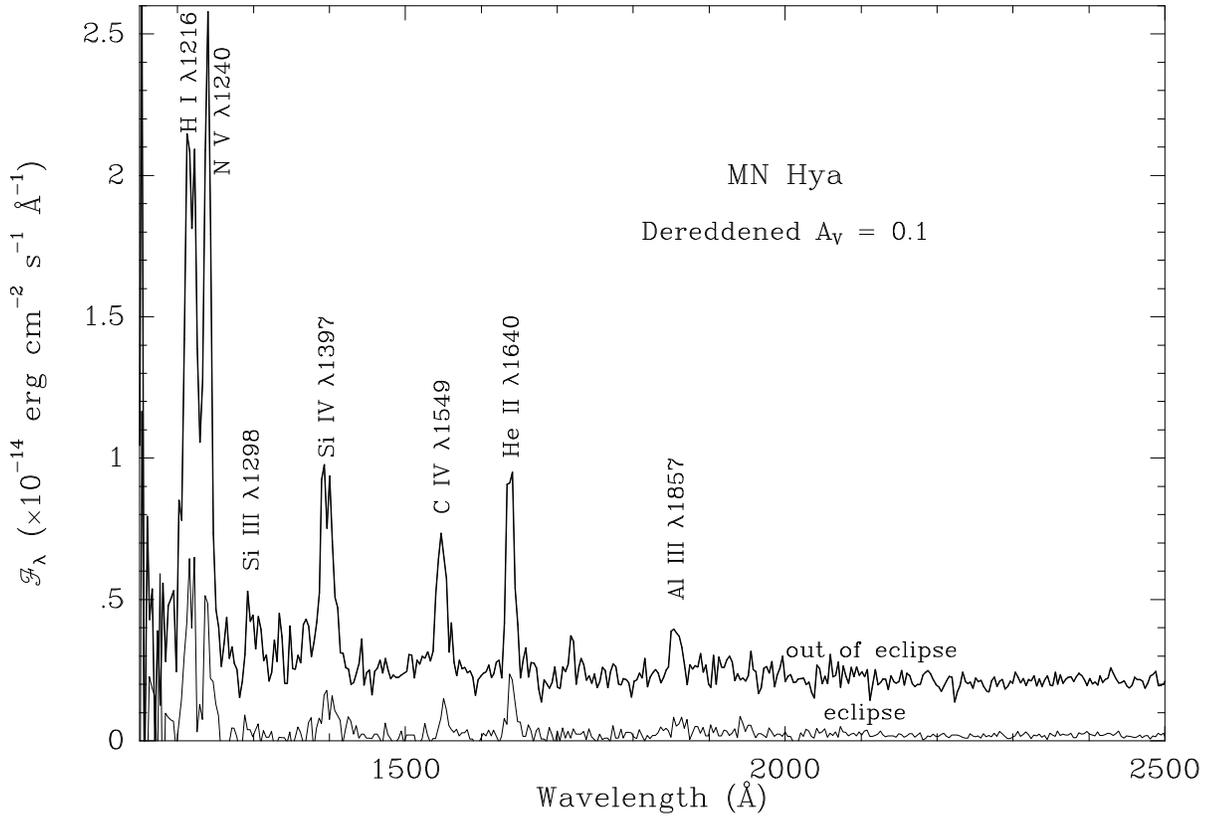}
\vskip1.truein
\caption{Dereddened in- and out-of-eclipse spectra of \mnhya.
As for \vori, \ion{N}{5} $\lambda$1240 is unusually strong and \ion{N}{4}
$\lambda$1718 is probably present. }
\end{figure}

\clearpage

\begin{figure}%Figure 10
\includegraphics[bb=40 670 2 2, scale=.9, angle=0]{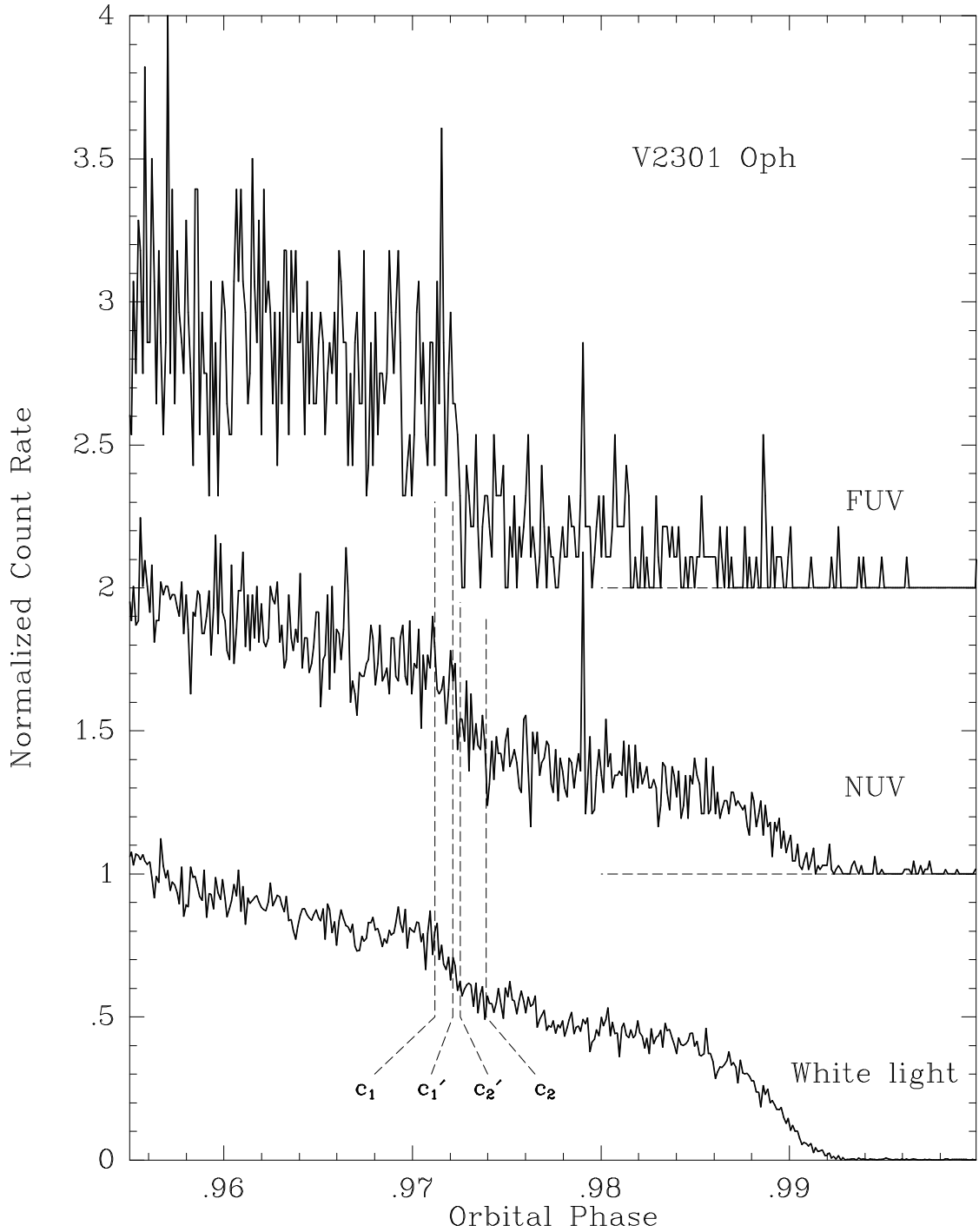}
\vskip7.3truein
\caption{Eclipse ingress of \voph.  Two pairs of ingress
contacts are indicated by the data: $c_1,c_2$ in the NUV channel, suggesting a
structure approximately the size of the white dwarf, and
$c_1\arcmin,c_2\arcmin$ in the FUV, which is evidence for an accretion-heated
spot on the white dwarf of linear dimension $\sim$$1.4\times10^8$~cm and
$f_{\rm spot}\sim0.008$. The extended decline to minimum light in all bands
corresponds to the eclipse of the portion of the accretion stream which
overshoots in azimuth the white dwarf.}
\end{figure}

\clearpage

\begin{figure}%Figure 11
\includegraphics[bb=420 50 2 2, scale=.7, angle=-90]{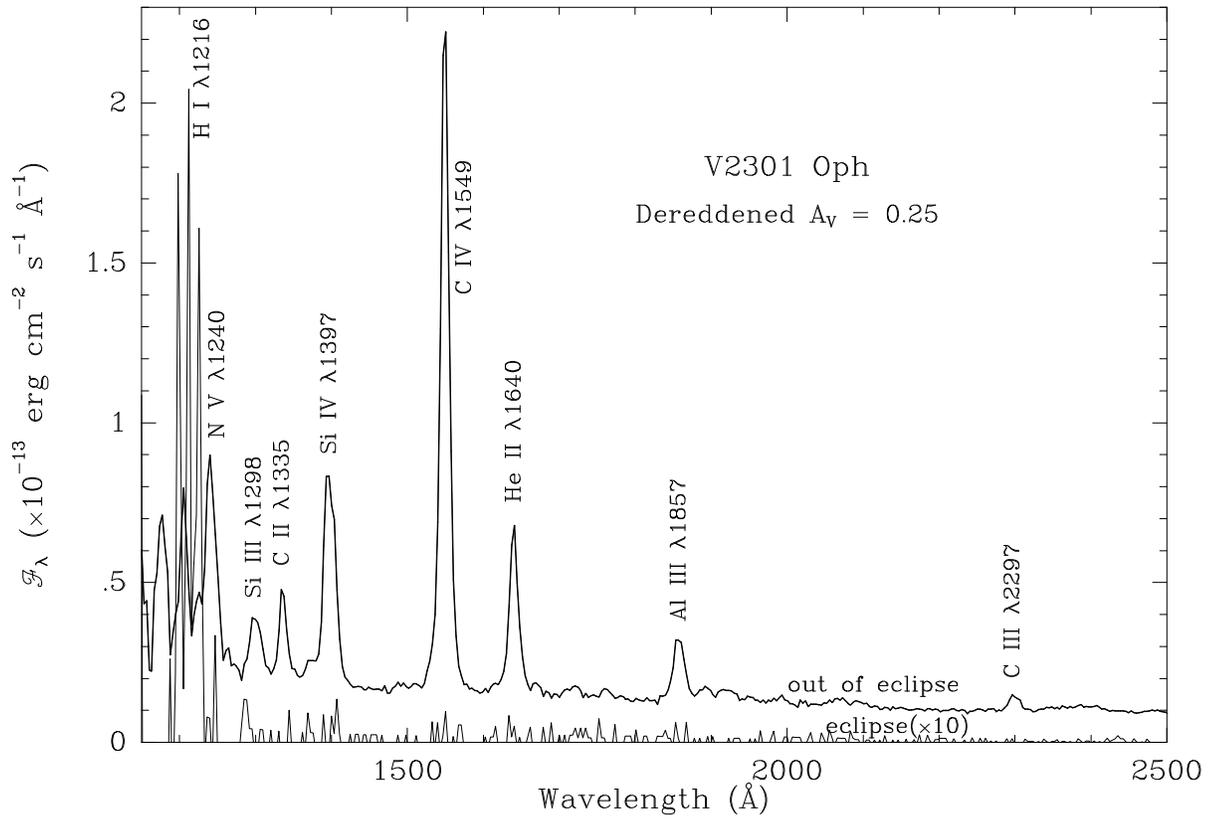}
\vskip1.truein
\caption{Dereddened in- and out-of-eclipse spectra of \voph.}
\end{figure}

\clearpage

\begin{figure}%Figure 12
\includegraphics[bb=40 610 2 2, scale=.85, angle=0]{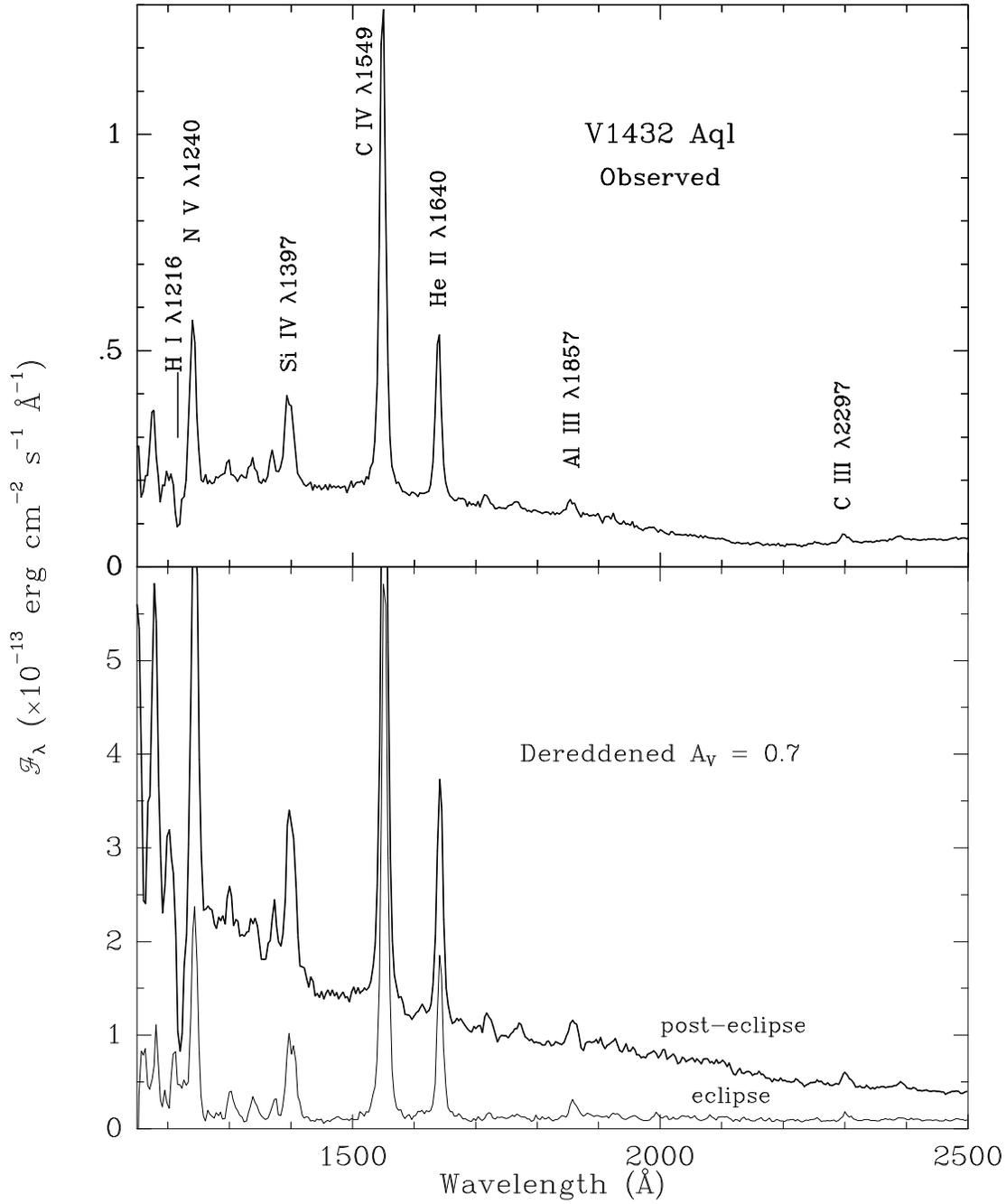}
\vskip6.2truein \caption{FOS spectra of \vaql.  {\it (Top):\/} The mean
spectrum over all epochs, as observed.  {\it (Bottom:)\/} Eclipse and
post-eclipse spectra taken from the datasets of 1996 Aug. 29, corrected for an
interstellar extinction of $A_v=0.7$~mag. Note the steep continuum out of
eclipse.}
\end{figure}

\end{document}